\documentclass[aps,pra,showpacs,superscriptaddress,preprint]{revtex4}
\usepackage{amsmath}
\usepackage{graphicx}

\catcode`ð=\active
 \defð{\u{g}}
 \catcode`Ð=\active
 \defÐ{\u{G}}
 \catcode`Ý=\active
\defÝ{\. I}
 \catcode`ö=\active
\defö{\"{o}}
 \catcode`Ö=\active
 \defÖ{\"O}
 \catcode`ü=\active
 \defü{\"{u}}
 \catcode`Ü=\active
 \defÜ{\"{U}}
 \catcode`Þ=\active
\defÞ{\c{S}}
 \catcode`þ=\active
 \defþ{\c{s}}
 \catcode`ý=\active
 \defý{{\i}}
 \catcode`ç=\active
\defç{\d{c}}
 \catcode`Ç=\active
\defÇ{\d{C}}

\input{tcilatex}

\begin{document}

\title{Approximate $\kappa $-state solutions to the Dirac-Yukawa problem
based on the spin and pseudospin symmetry }
\author{Sameer M. Ikhdair}
\email[E-mail: ]{sikhdair@neu.edu.tr}
\affiliation{Physics Department, Near East University, Nicosia, North Cyprus, Turkey}
\date{%
\today%
}

\begin{abstract}
Using an approximation scheme to deal with the centrifugal
(pseudo-centrifugal) term, we solve the Dirac equation with the screened
Coulomb (Yukawa) potential for any arbitrary spin-orbit quantum number $%
\kappa .$ Based on the spin and pseudospin symmetry, analytic bound state
energy spectrum formulas and their corresponding upper- and lower-spinor
components of two Dirac particles are obtained using a shortcut of the
Nikiforov-Uvarov method. We find a wide range of permissible values for the
spin symmetry constant $C_{s}$ from the valence energy spectrum of particle
and also for pseudospin symmetry constant $C_{ps}$ from the hole energy
spectrum of antiparticle. Further, we show that the present potential
interaction becomes less (more) attractive for a long (short) range
screening parameter $\alpha $. To remove the degeneracies in energy levels
we consider the spin and pseudospin solution of Dirac equation for Yukawa
potential plus a centrifugal-like term. A few special cases such as the
exact spin (pseudospin) symmetry Dirac-Yukawa, the Yukawa plus
centrifugal-like potentials, the limit when $\alpha $ becomes zero (Coulomb
potential field) and the non-relativistic limit of our solution are studied.
The nonrelativistic solutions are compared with those obtained by other
methods.

\end{abstract}

\maketitle

\section{Introduction}

The screened Coulomb (Yukawa) potential is widely used in physics, being a
good approximation to short-range interactions between charged particles in
various areas of physics [1,2]. In plasma physics it is known as the Debye-H%
\"{u}ckel potential describes the shielding effect of ions embedded in
plasmas [3]. It has also been used to play a fundamental role in
(dusty/complex) plasma and colloidal suspensions. The momentum transfer in
pair collisions of particles interacting via the Yukawa potential is well
investigated in the limit when the interaction is \textquotedblleft
weak\textquotedblright\ in the sense that its range (distance at which the
interaction energy is equal to the kinetic energy) is much shorter than the
plasma screening length. This limit is known as the theory of Coulomb
scattering and is extensively used to describe collisions in usual
electron-ion plasma [4]. In solid state, atomic and molecular physics it is
called the Thomas-Fermi or screened Coulomb potential due to the cloud of
electronic charges around the nucleus [5,6]. Also this potential is well
known in nuclear physics as the dominant central part of neutrons-protons
nuclear interaction due to the massive field exchange (one pion) whose mass
is $m$ [7,8]. In high energy physics, the potential is used to model the
interaction of hadrons in short range gauge theories where coupling is
mediated by the exchange of a massive scalar meson [1,9]. It is defined as
follows [10]: 
\begin{equation}
V(r)=-\frac{A}{r}e^{-\alpha r},\text{ }A>0
\end{equation}%
where $\alpha $ and $A$ are the screening (range) and coupling strength
parameters, respectively. The two parameters are given by different
expressions depending on the type of the problem under consideration. For
example, $A=g^{2}$ is positive for attraction$,$ $g$ denotes the coupling
constant between meson field and the fermion field with which it interacts.
Since the field mediator is massive, the corresponding force has a certain
range, which is inversely proportional to the mass, $\alpha =mc/\hbar $ \ If
the mass is zero, then the Yukawa potential becomes equal to a Coulomb
potential and the range is said to be infinite. Further, the number of bound
states of the Yukawa potential is found to be finite. Unfortunately, since
the Schr\"{o}dinger equation for the screened Coulomb potential does not
admit an exact analytical solution [11], therefore, various numerical
[12,13] and analytical [14-17] methods have been developed in the past. Also
the energy spectrum can be calculated with high accuracy by means of the
hypervirial relations and Pade approximation methods [18,19]. The
short-range behavior of the decaying exponential factor $e^{-\alpha r}$ and
singularity at $r=0$ make the task of obtaining accurate solutions a
difficult task. Besides, most of these calculations suffer from limited
accuracy when a wider range of potential parameters are being considered
[20].

An approximate perturbative method has been developed to obtain the energy
spectrum and wave functions of the Schr\"{o}dinger with the Yukawa-like
potentials [21]. This method has been applied to obtain the energy spectrum
and wave functions for the more general exponential-cosine-screened Coulomb
potentials. These potentials are containing an additional Coulomb term
superposed with the Yukawa potential that might be useful in describing the
effective interaction in many-body problem [22]. Further, the asymptotic
iteration method is used to obtain the energy eigenvalues of the Yukawa
potential [23]. The $J$-matrix method has been applied to the Yukawa
potential with no special treatment of its singularity by using the
oscillator basis and the reference Hamiltonian contained only the kinetic
energy operator [24]. The bound states spectrum and resonance energies of
the Yukawa potential have been studied using the method of complex scaling
[25]. Therefore, an alternative Laguerre basis has also been used to comply
with the $J$-matrix requirement of a tridiagonal matrix representation of
the reference Hamiltonian that includes the $r^{-1}$ singularity [18].

When a particle is in a strong interaction (range of interaction exceeds the
screening length, $\lambda =\alpha ^{-1}$), the relativistic effect must be
considered which gives the correction for nonrelativistic quantum mechanics.
The relativistic treatment is of much interest especially when (at least)
one of the particles is highly charged and their relative velocity is small.
Taking the relativistic effects into account, a particle in a potential
field should be described with the Dirac equation. Therefore, the solution
of the Dirac equation can be important in different fields of physics like
nuclear and molecular physics [7,26]. Within the framework of the Dirac
equation the spin symmetry arises if the magnitude of the spherical
attractive scalar potential $S(r)$ and repulsive vector potential are nearly
equal (i.e., $S(r)\sim V(r))$ in the nuclei (\textit{i.e.}, when the
difference potential $\Delta (r)=V(r)-S(r)=C_{s},$ with $C_{s}$ is an
arbitrary constant$);$ however, the pseudospin symmetry occurs if $S(r)\sim
-V(r)$ are nearly equal (\textit{i.e.}, when the sum potential $\Sigma
(r)=V(r)+S(r)=C_{ps},$ with $C_{ps}$ is an arbitrary constant$)$ [27]$.$ The
spin symmetry is relevant for mesons [28]. The pseudospin symmetry concept
has been applied to many systems in nuclear physics and related areas
[27-31] and used to explain features of deformed nuclei [32], the
super-deformation [33] and to establish an effective nuclear shell-model
scheme [29,30,34]. Recently, the spin and pseudospin symmetries have been
widely applied on several physical potentials by many authors (cf. [35-37]
and references therein). Many authors have investigated approximately the
solution of the Dirac equation with a few potential models such as the the
generalized Morse potential [35], the Hulth$\mathbf{{\acute{e}}}$n potential
[36], the Rosen-Morse potential [37] and the screened Coulomb potential [38]
etc within the framework of various methods.

In the framework of the spin symmetry $S(r)\sim V(r)$ and pseudospin
symmetry $S(r)\sim -V(r)$, the bound state energy eigenvalues and associated
upper- and lower-spinor wave functions are investigated by means of the
Nikiforov-Uvarov (NU) method [39]. We have approximately solved the Dirac
equation for the Hulthen potential [36] with spin and pseudospin symmetry
for any spin-orbit $\kappa $ state and found the eigenvalue equation and
corresponding two-component spinors within the framework of an approximation
to the term proportional to $1/r^{2}.$ We have also solved the (3+1)
dimensional Dirac equation for a single particle trapped in the spherically
symmetric generalized WS potential under the conditions of exact spin and
pseudospin symmetry combined with approximation for the spin-orbit
centrifugal (pseudo-centrifugal) term, and calculated the two-component
spinor wave functions and the energy eigenvalues for any arbitrary
spin-orbit $\kappa $ bound states [40]. Recently, Setare and Haidari [41]
have solved the Dirac-Yukawa problem in the presence of the spin symmetry
and given only analytical expressions for energy eigenvalues and wave
functions. However, they have not given further numerical discussions for
the validity of their analytical solutions. On the other hand, the subject
of the pseudospin symmetry of the Dirac-Yukawa problem introduced by
Ginocchio [27] has not been investigated by Ref. [41]. Over the past years,
the interest in the quality of the pseudospin symmetry has been increased in
the framework of the single-particle relativistic potential models.
Therefore, we have found that it is necessary to give a detailed study for
the solution of the Dirac equation with screened Coulomb (Yukawa) potential
model in the presence of spin as well as pseudospin symmetry in a very
simple and elegant way by using a shortcut procedures for the NU method. We
also give a detailed discussion for the validity of the present numerical as
well as analytical solutions. We also try to explore the exact relativistic
energy spectrum of the Coulombic field (when $\alpha =0,$ the low screening
range of the Yukawa potential) under the exact spin and pseudospin symmetry.

The analytic solution of the Dirac equation with the screened Coulomb
potential is difficult to find due to the centrifugal (pseudo centrifugal)
term $\kappa (\kappa +1)r^{-2}$ ($\kappa (\kappa -1)r^{-2}$) and the
singular interactions like $r^{-1}$ (e.g., the Coulomb potential).
Nevertheless, employing the approximation provided by Greene and Aldrich
[42] to the centrifugal term $\ r^{-2}$ and to the singular Coulombic part $%
\ r^{-1}$ makes the solution handy$.$ We work within the framework of the
low screening parameter throughout the paper. We find analytically
approximate bound state solutions including the energy spectra and the
corresponding spinor wave functions in the presence of the spin symmetry and
pseudospin symmetry concept for any $\kappa $-state within the parametric
generalization of the NU method [43] given in Appendix A

This paper is organized as follows. In section 2, we investigate the bound
state energy equation and the corresponding two-component spinor wave
functions in the presence of spin and pseudospin symmetry concept for the
screened Coulomb potential by employing a parametric generalization of the
NU method. In section 3, we study some special cases like Schr\"{o}%
dinger-Yukawa, Dirac-Coulomb, exact spin (pseudospin) symmetric Dirac-Yukawa
problem and Yukawa plus centrifugal-like potentials. In section 4, we
present some numerical results to the non-relativistic and relativistic
numerical energy levels for the Yukawa potential. The relevant conclusions
are given in section 5.

\section{Theoretical Framework of Dirac bound state solutions}

The Dirac equation for fermionic massive spin-$1/2$ particles moving in an
attractive scalar potential $S(r)$ and a repulsive vector potential $V(r)$
is given by [7] 
\begin{equation}
\left[ c\mathbf{\alpha }\cdot \mathbf{p+\beta }\left( Mc^{2}+S(r)\right)
+V(r)-E\right] \psi _{n\kappa }(\mathbf{r})=0,\text{ }\psi _{n\kappa }(%
\mathbf{r})=\psi (r,\theta ,\phi ),
\end{equation}%
where $E$ is the relativistic energy of the system, $M$ is the mass of a
particle, $\mathbf{p}=-i\hbar \mathbf{\nabla }$ is the momentum operator,
and $\mathbf{\alpha }$ and $\mathbf{\beta }$ are $4\times 4$ Dirac matrices,
i.e.,%
\begin{equation}
\mathbf{\alpha =}\left( 
\begin{array}{cc}
0 & \mathbf{\sigma }_{i} \\ 
\mathbf{\sigma }_{i} & 0%
\end{array}%
\right) ,\text{ }\mathbf{\beta =}\left( 
\begin{array}{cc}
\mathbf{I} & 0 \\ 
0 & -\mathbf{I}%
\end{array}%
\right) ,\text{ }\sigma _{1}\mathbf{=}\left( 
\begin{array}{cc}
0 & 1 \\ 
1 & 0%
\end{array}%
\right) ,\text{ }\sigma _{2}\mathbf{=}\left( 
\begin{array}{cc}
0 & -i \\ 
i & 0%
\end{array}%
\right) ,\text{ }\sigma _{3}\mathbf{=}\left( 
\begin{array}{cc}
1 & 0 \\ 
0 & -1%
\end{array}%
\right) ,
\end{equation}%
where $\mathbf{I}$ denotes the $2\times 2$ identity matrix and $\mathbf{%
\sigma }_{i}$ are the three-vector Pauli spin matrices. In a spherical
symmetrical nuclei, the total angular momentum operator of the nuclei $%
\mathbf{J}$ and spin-orbit matrix operator $\mathbf{K}=-\mathbf{\beta }%
\left( \mathbf{\sigma }\cdot \mathbf{L}+\mathbf{I}\right) $ commute with the
Dirac Hamiltonian, where $\mathbf{L}$ is the orbital angular momentum
operator. The spinor wave functions can be classified according to the
radial quantum number $n$ and the spin-orbit quantum number $\kappa $ and
can be written using the Pauli-Dirac representation in the following forms:%
\begin{equation}
\psi _{n\kappa }(\mathbf{r})=\left( 
\begin{array}{c}
f_{n\kappa }(\mathbf{r}) \\ 
g_{n\kappa }(\mathbf{r})%
\end{array}%
\right) =\frac{1}{r}\left( 
\begin{array}{c}
F_{n\kappa }(r)Y_{jm\kappa }^{l}(\widehat{r}) \\ 
iG_{n\kappa }(r)Y_{jm(-\kappa )}^{\widetilde{l}}(\widehat{r})%
\end{array}%
\right) ,
\end{equation}%
where the upper- and lower-spinor components $F_{n\kappa }(r)$ and $%
G_{n\kappa }(r)$ are the real square-integral radial wave functions, $%
Y_{jm\kappa }^{l}(\widehat{r})$ and $Y_{jm(-\kappa )}^{\widetilde{l}}(%
\widehat{r})$ are the spin spherical harmonic functions coupled to the total
angular momentum $j$ and it's projection $m$ on the $z$ axis and $\kappa
\left( \kappa +1\right) =l(l+1)$ and $\kappa \left( \kappa -1\right) =%
\widetilde{l}(\widetilde{l}+1)$. The quantum number $\kappa $ is related to
the quantum numbers for spin symmetry $l$ and pseudospin symmetry $%
\widetilde{l}$ as%
\begin{equation}
\kappa =\left\{ 
\begin{array}{cccc}
-\left( l+1\right) =-\left( j+\frac{1}{2}\right) , & (s_{1/2},p_{3/2},\text{%
\textit{etc.}}), & \text{ }j=l+\frac{1}{2}, & \text{aligned spin }\left(
\kappa <0\right) , \\ 
+l=+\left( j+\frac{1}{2}\right) , & \text{ }(p_{1/2},d_{3/2},\text{\textit{%
etc.}}), & \text{ }j=l-\frac{1}{2}, & \text{unaligned spin }\left( \kappa
>0\right) ,%
\end{array}%
\right.
\end{equation}%
and the quasi-degenerate doublet structure can be expressed in terms of a
pseudospin angular momentum $\widetilde{s}=1/2$ and pseudo-orbital angular
momentum $\widetilde{l}$ which is defined as 
\begin{equation}
\kappa =\left\{ 
\begin{array}{cccc}
-\widetilde{l}=-\left( j+\frac{1}{2}\right) , & (s_{1/2},\text{ }p_{3/2},%
\text{ \textit{etc.}}), & j=\widetilde{l}-1/2, & \text{aligned spin }\left(
\kappa <0\right) , \\ 
+\left( \widetilde{l}+1\right) =+\left( j+\frac{1}{2}\right) , & \text{ }%
(d_{3/2},\text{ }f_{5/2},\text{ \textit{etc.}}), & \ j=\widetilde{l}+1/2, & 
\text{unaligned spin }\left( \kappa >0\right) ,%
\end{array}%
\right.
\end{equation}%
where $\kappa =\pm 1,\pm 2,\cdots .$ For example, ($1s_{1/2},0d_{3/2}$) and
(2p$_{3/2},1f_{5/2}$) can be considered as pseudospin doublets.

Upon direct substitution of Eq. (4) into Eq. (2), the two radial coupled
Dirac equations for the two spinor components can be obtained as 
\begin{subequations}
\begin{equation}
\left( \frac{d}{dr}+\frac{\kappa }{r}\right) F_{n\kappa }(r)=\left[
Mc^{2}+E_{n\kappa }-\Delta (r)\right] G_{n\kappa }(r),
\end{equation}%
\begin{equation}
\left( \frac{d}{dr}-\frac{\kappa }{r}\right) G_{n\kappa }(r)=\left[
Mc^{2}-E_{n\kappa }+\Sigma (r)\right] F_{n\kappa }(r),
\end{equation}%
where $\Delta (r)=V(r)-S(r)$ and $\Sigma (r)=V(r)+S(r)$ are the difference
and sum potentials, respectively.

In the presence of the spin symmetry ( i.e., $\Delta (r)=C_{s}=$ constant),
one can eliminate $G_{n\kappa }(r)$ in Eq. (7a), with the aid of Eq. (7b),
to obtain a second-order differential equation for the upper-spinor
component: 
\end{subequations}
\begin{equation*}
\left[ -\frac{d^{2}}{dr^{2}}+\frac{\kappa \left( \kappa +1\right) }{r^{2}}+%
\frac{1}{\hbar ^{2}c^{2}}\left( E_{n\kappa }+Mc^{2}-C_{s}\right) \Sigma (r)%
\right] F_{n\kappa }(r)
\end{equation*}%
\begin{equation}
=\frac{1}{\hbar ^{2}c^{2}}\left[ E_{n\kappa }^{2}-M^{2}c^{4}-C_{s}\left(
E_{n\kappa }-Mc^{2}\right) \right] F_{n\kappa }(r),
\end{equation}%
and the lower-spinor component is obtained from Eq. (7a):%
\begin{equation}
G_{n\kappa }(r)=\frac{1}{Mc^{2}+E_{n\kappa }-C_{s}}\left( \frac{d}{dr}+\frac{%
\kappa }{r}\right) F_{n\kappa }(r),
\end{equation}%
where $E_{n\kappa }\neq -Mc^{2},$ only real positive energy spectrum exist
when $C_{s}=0$ (exact spin symmetry). On the other hand, in the presence of
the pseudospin symmetry ( i.e., $\Sigma (r)=C_{ps}=$ constant), one can
eliminate $F_{n\kappa }(r)$ in Eq. (7b), with the aid of Eq. (7a), to obtain
a second-order differential equation for the lower-spinor component:%
\begin{equation*}
\left[ -\frac{d^{2}}{dr^{2}}+\frac{\kappa \left( \kappa -1\right) }{r^{2}}-%
\frac{1}{\hbar ^{2}c^{2}}\left( Mc^{2}-E_{n\kappa }+C_{ps}\right) \Delta (r%
\right] G_{n\kappa }(r)
\end{equation*}%
\begin{equation}
=\frac{1}{\hbar ^{2}c^{2}}\left[ E_{n\kappa }^{2}-M^{2}c^{4}-C_{ps}\left(
E_{n\kappa }+Mc^{2}\right) \right] G_{n\kappa }(r),
\end{equation}%
and the upper-spinor component $F_{n\kappa }(r)$ can be obtained from Eq.
(7b) as%
\begin{equation}
F_{n\kappa }(r)=\frac{1}{Mc^{2}-E_{n\kappa }+C_{ps}}\left( \frac{d}{dr}-%
\frac{\kappa }{r}\right) G_{n\kappa }(r),
\end{equation}%
where $E_{n\kappa }\neq Mc^{2},$ only real negative energy spectrum exist
when $C_{ps}=0$ (exact pseudospin symmetry). The physical solution demands
that the upper and lower radial components should satisfy the boundary
conditions: $F_{n\kappa }(0)=G_{n\kappa }(0)=0$ and $F_{n\kappa }(\infty
)=G_{n\kappa }(\infty )=0.$

\subsection{Spin symmetry Dirac-Yukawa problem}

At first, we investigate the spin symmetry in the form of $SU(2)$ by taking
the $\Sigma (r)=2V(r)\rightarrow V_{Y}(r)$ [44] which can be easily reduced
into the non-relativistic limit under a certain appropriate transformations.
Equation (8) shows that the energy eigenvalues, $E_{n\kappa }$ is mainly
dependent on the quantum numbers $n$ and $l.$ For example, when $l\neq 0,$
the states with $j=l\pm 1/2$ are degenerate. The sum potential $\Sigma (r)$
in Eq. (8) is simply taken as the Yukawa potential,%
\begin{equation}
\Sigma (r)=-\frac{\Sigma _{0}}{r}e^{-\alpha r},\text{ }\Sigma _{0}=A>0,
\end{equation}%
which provides a simple Schr\"{o}dinger-like equation in the form:%
\begin{equation}
\left[ \frac{d^{2}}{dr^{2}}-\frac{\kappa \left( \kappa +1\right) }{r^{2}}%
-4\alpha ^{2}\nu _{1}^{2}+2\alpha \omega _{1}\frac{A}{r}e^{-\alpha r}\right]
F_{n\kappa }(r)=0,\text{ }\kappa \left( \kappa +1\right) =l\left( l+1\right)
\end{equation}%
where%
\begin{equation}
\nu _{1}^{2}=\frac{1}{4\alpha ^{2}\hbar ^{2}c^{2}}\left( Mc^{2}-E_{n\kappa
}\right) \left( Mc^{2}+E_{n\kappa }-C_{s}\right) ,\text{ \ }\omega _{1}=%
\frac{1}{2\alpha \hbar ^{2}c^{2}}\left[ E_{n\kappa }+Mc^{2}-C_{s}\right] A,
\end{equation}%
with $\kappa $ values are given in Eq. (5). The exact analytic solution of
Eq. (13) is difficult to find due to the centrifugal kinetic energy term $%
\kappa (\kappa +1)r^{-2}$ and the singularity of $r^{-1}$-type$.$
Nonetheless, if $\kappa $ is not too large, the case of the vibrations of
small amplitude about the minimum, we attempt to use the Greene-Aldrich [42]
conventional approximation to deal with centrifugal term, 
\begin{equation}
\frac{1}{r^{2}}\approx 4\alpha ^{2}\frac{e^{-2\alpha r}}{\left(
1-e^{-2\alpha r}\right) ^{2}}.
\end{equation}%
Introducing the new parameter, $x(r)=e^{-2\alpha r}\in \lbrack 0,1]$\ and
further substituting Eq. (15) into Eq. (13), we obtain 
\begin{equation*}
\left\{ \frac{d^{2}}{dx^{2}}+\frac{(1-x)}{x(1-x)}\frac{d}{dx}\right.
\end{equation*}%
\begin{equation}
+\left. \frac{\left[ -\left( \nu _{1}^{2}+\omega _{1}\right) x^{2}+\left(
2\nu _{1}^{2}+\omega _{1}-\kappa \left( \kappa +1\right) \right) x-\nu
_{1}^{2}\right] }{x^{2}(1-x)^{2}}\right\} F_{n\kappa }(x)=0,
\end{equation}%
where we have inserted $F_{n\kappa }(r)=F_{n\kappa }(x).$ In order to
clarify the parametric generalization of the NU method [35,37,43], let us
take the following general form of a Schr\"{o}dinger-like equation written
for any potential, 
\begin{equation}
\left[ \frac{d^{2}}{dx^{2}}+\frac{\widetilde{\tau }(x)}{\sigma (x)}\frac{d}{%
dx}+\frac{\widetilde{\sigma }(x)}{\sigma ^{2}(x)}\right] \psi _{n}(x)=0,
\end{equation}%
satisfying the wave functions%
\begin{equation}
\psi _{n}(x)=\phi (x)y_{n}(x).
\end{equation}%
In addition, the two polynomials 
\begin{equation}
\widetilde{\tau }(x)=c_{1}-c_{2}x,
\end{equation}%
and 
\begin{equation}
\sigma (x)=x\left( 1-c_{3}x\right) \text{ and \ }\widetilde{\sigma }(x)=-\xi
_{2}x^{2}+\xi _{1}x-\xi _{0},
\end{equation}%
are at most of first- and second-degree, respectively. Comparing Eq. (17)
with its counterpart Eq. (16), we obtain values for the constants $c_{i}$ ($%
i=1,$2$,3)$ along with $\xi _{j}$ ($j=0,1,2).$ Now, following the NU method
[39] and making the substitution of Eqs. (19) and (20) leads to more general
forms for the polynomials $\pi (z)$ and $\tau (z),$ the root $k,$ the
eigenvalues equation and the wave functions $\phi (z)$ and $y_{n}(z)$ all
expressed in terms of the constants $c_{i}$ ($i=4,$ $5,\cdots ,13)$ as given
in Appendix A$.$ Hence, the task of computing the energy eigenvalues and the
corresponding wave functions of Eq. (13) within the framework of the
parametric generalization of the NU method becomes relatively easy and
straightforward. It may be explained shortly in the following steps:

Firstly, we need to find the specific values for the parametric constants $%
c_{i}$ ($i=4,$ $5,\cdots ,13)$ by means of the relation A1 of Appendix A.
The values of all these constants $c_{i}$ ($i=1,$ $2,\cdots ,13)$ together
with $\xi _{j}$ ($j=0,1,2)$ are therefore listed in Table 1 for the screened
Coulomb potential model.

Secondly, by using the relations (A2-A5), the analytic forms of the
essential polynomials $\pi (x)$ and $\tau (x)$ along with the root $k,$
required by the NU method, can also be found as%
\begin{equation}
\pi (x)=-\frac{x}{2}-\frac{1}{2}\left[ \left( 2\kappa +1+2\nu _{1}\right)
x-2\nu _{1}\right] ,
\end{equation}%
\begin{equation}
k=-\kappa \left( \kappa +1\right) +\omega _{1}-\left( 2\kappa +1\right) \nu
_{1},
\end{equation}%
and%
\begin{equation}
\tau (x)=1+2\nu _{1}-2\left( 1+\nu _{1}+\frac{1}{2}\left( 2\kappa +1\right)
\right) x,
\end{equation}%
where $\tau ^{\prime }(x)<0$ must be satisfied in order to obtain physical
solution according to the NU method [39]$.$

Thirdly, we need to calculate the energy eigenvalues by means of the
eigenvalue equation, relation A6 which gives%
\begin{equation}
\left( n+\kappa +1\right) ^{2}+2\left( n+\kappa +1\right) \nu _{1}=\omega
_{1}.
\end{equation}%
Finally, after making use of Eq. (14), the above equation for the Yukawa
potential can be expressed implicitly in terms of the energy $E_{n\kappa }$
as 
\begin{equation*}
\sqrt{\left( Mc^{2}-E_{n\kappa }\right) \left( Mc^{2}+E_{n\kappa
}-C_{s}\right) }+\alpha \hbar c\left( n+\kappa +1\right) =\frac{\left(
Mc^{2}+E_{n\kappa }-C_{s}\right) AZe^{2}}{2\hbar c\left( n+\kappa +1\right) }%
,
\end{equation*}%
\begin{equation}
Mc^{2}>E_{n\kappa }\text{ and }Mc^{2}+E_{n\kappa }>C_{s},\text{ }%
n=0,1,2,\cdots ,
\end{equation}%
where $\kappa \neq -(n+1)$ and $A=Ze^{2}.$ The above energy equation can be
also rearranged in a quadratic form (in relativistic units $\hbar =c=1$) as%
\begin{equation}
\left[ 1+\left( \frac{A}{N_{1}}\right) ^{2}\right] E_{n\kappa }^{2}-2\left[
S+\left( \frac{A}{N_{1}}\right) ^{2}W\right] E_{n\kappa }+\left( \frac{AW}{%
N_{1}}\right) ^{2}+\left( \frac{\alpha N_{1}}{2}\right) ^{2}+\left( M+\alpha
A\right) W=0,
\end{equation}%
where 
\begin{equation}
N_{1}=2\left( n+\kappa +1\right) ,\text{ }W=C_{s}-M,\text{ }S=\frac{1}{2}%
(C_{s}+\alpha A).
\end{equation}%
The two energy spectrum formula of the quadratic equation (26) is 
\begin{equation}
E_{n\kappa }^{\pm }=\frac{(A^{2}W+SN_{1}^{2})\pm \sqrt{%
(A^{2}W+SN_{1}^{2})^{2}-(A^{2}+N_{1}^{2})\left[ \left( AW+\frac{1}{2}\alpha
N_{1}^{2}\right) ^{2}+MWN_{1}^{2}\right] }}{A^{2}+N_{1}^{2}},
\end{equation}%
where $(A^{2}W+SN_{1}^{2})\geq \sqrt{\left( A^{2}+N_{1}^{2}\right) \left[
\left( AW+\frac{1}{2}\alpha N_{1}^{2}\right) ^{2}+MWN_{1}^{2}\right] }$ for
distinct particle and anti-particle real energy bound states $E_{n\kappa
}^{p}=E_{n\kappa }^{+}$ and $E_{n\kappa }^{a}=E_{n\kappa }^{-},$
respectively. Otherwise, in the case when the above inequality does not
hold, we will have no bound state solutions (scattering states).

Figure 1 shows the ground state valence energy level of particle and
antiparticle as a function of different values of the coupling constant $A$
and the screening parameter (range) of the potential $\alpha =0.01,0.02,0.05$
and $0.10$ for two degenerate partners ($n=0,\kappa =1$) and ($n=3,\kappa
=-2 $) labelled as ($0p_{1/2},3p_{3/2}$) of particle and antiparticle with
the parameters choices of mass $M=5.0$ $fm^{-1}$ and spin constant $%
C_{s}=4.9 $ $fm^{-1},$ as it should be expected, for a given value of $%
\alpha $ the bound state becomes sharply (slowly) deeper (more attractive)
for particle (antiparticle) on increasing the coupling constant $A$ (heavy
nucleus). On the decreasing the value of the range $\alpha $ (lower
screening range), the energy goes to a more negative value for antiparticle
and to a less positive value for particle (energy becomes more attractive)$.$
On the other hand, for a fixed value of $A,$ the bound state energy becomes
shallower on increasing $\alpha $ for particle. It is seen that in the limit
of a very short-ranged potential ($\alpha \rightarrow 0$), the potential
approaches the $\delta $-function limit that can bind particles and
antiparticle stronger than finite-ranged potentials (1). Furthermore, Fig. 2
shows the variation of ground state valence energy level of particle and
antiparticle with the spin constant $C_{s}$ for several values of the
spin-orbit $\kappa =1,3$ and $5$ with special choices of parameters $M=5.0$ $%
fm^{-1},$ $\alpha =0.1$ $fm^{-1}$ and $A=1.$ A very careful inspection for
both numerical results and Fig. 2a shows that there is a small energy
difference between the states $\kappa =1,3$ and $5$ although the values of
spin constant $C_{s}$ increases in the range $11.4$ $fm^{-1}\leq C_{s}\leq
20 $ $fm^{-1},$ i.e., the energy spectrum is not sensitive to the influence
of $\kappa $ in the aforementioned range. With the increasing $\kappa $ value%
$,$ we see that $E_{0\kappa }^{+}$ fan out toward the stronger positive
energy spectrum for the given range $-20$ $fm^{-1}\leq C_{s}\leq 8.8$ $%
fm^{-1}.$ The scattering (not bound) states could be seen in the range $8.8$ 
$fm^{-1}\leq C_{s}\leq 11.3$ $fm^{-1}$ for $\kappa =1,3$ and $5.$ The
physical explanation to Figure 2a is being illustrated as follows. It is
seen in Table 2 that the energy spectrum (positive and negative) of Eq. (25)
is entirely dependent on the choice of $C_{s},$ i.e., $E_{0\kappa
}^{+}=E_{0\kappa }^{+}(C_{s}).$ For example, in the single electron
interaction with nucleus $Ze$ (units $\hbar =c=e=1),$ we may have two
essential requirements: (a) $M\geq E_{n\kappa }$ and $M+E_{n\kappa }\geq
C_{s},$ in which both impose a restriction on the choice of the range values
of the permissible spin symmetry constant $C_{s}$ are in the interval $-20$ $%
fm^{-1}\leq C_{s}\leq 8.8$ $fm^{-1}$ with a requirement that energy spectrum
be real along with $\left\vert E_{n\kappa }\right\vert <M$ and (b) $M\leq
E_{n\kappa }$ and $M+E_{n\kappa }\leq C_{s}$ which is possible for\ $11.4$ $%
fm^{-1}\leq C_{s}\leq 20$ $fm^{-1}$ where $E_{n\kappa }>M.$ In the two cases
we have taken the spin-orbit quantum number $\kappa =1,3$ and $5.$ While
taking $\kappa =1,3,5,7$ and $9$, $C_{s}$ lies in the two intervals: $-20$ $%
fm^{-1}\leq C_{s}\leq 8.0$ $fm^{-1}$ and $12.2$ $fm^{-1}\leq C_{s}\leq 20$ $%
fm^{-1}.$ Nonetheless, it is necessary to choose the physical solution as
shown in case (a) since it is consistent with the exact spin symmetry when $%
C_{s}=0$ (i.e., $S(r)=V(r)$) $(\left\vert E_{n\kappa }\right\vert <M)$ in
which the energy levels become very sensitive to the influence of $\kappa $
as usually expected$.$ Table 2 supports our choice of the allowed range $-20$
$fm^{-1}\leq C_{s}\leq 8.8$ $fm^{-1}.$ Moreover, we find out that the second
case is not sensitive to the influence of $\kappa .$ Thus, for our choice of 
$C_{s}=4.9$ $fm^{-1}$ which is falling in (a)$,$ we present all ground
energy spectrum including four states: $E_{0,1}=4.78641$ $fm^{-1},$ $%
E_{0,2}=4.94209$ $fm^{-1},$ $E_{0,3}=4.98904$ $fm^{-1}$ and $E_{0,4}=4.99998$
$fm^{-1}.$\ 

The results presented in Fig. 2b and Table 2 show that the energy difference
of antiparticle between the states $\kappa =1,3$ and $5$ shows a slight
change although the values of spin constant $C_{s}$ increases in the
interval $\ 10.6$ $fm^{-1}\leq C_{s}\leq 20$ $fm^{-1}$ (same as in Fig. 2a).
This range is forbidden since $E_{0,\kappa }^{-}<-M.$ As $\kappa $
decreasing, $E_{0\kappa }^{-}$ fan out toward the stronger (deeper) negative
energy spectrum when $-16$ $fm^{-1}\leq C_{s}\leq -0.1$ $fm^{-1}$ ($%
E_{0,\kappa }^{-}$ is very strongly negative, forbidden)$.$ In the intervals 
$0$ $fm^{-1}\leq C_{s}\leq 4.9$ $fm^{-1},$ $E_{0,\kappa }^{-}<M$ (negative,
permissible), $5.0$ $fm^{-1}\leq C_{s}\leq 9.6$ $fm^{-1},$ $E_{0,\kappa
}^{-}<M$ (positive, permissible) while $9.7$ $fm^{-1}\leq C_{s}\leq 10.6$ $%
fm^{-1}$ (energy is complex). The variation of the ground negative energy
level with $C_{s}$ contains forbidden values of $C_{s}$ that gives $%
E_{0,\kappa }\neq -M$ which results in an infinity wave function. The range
values of the allowed $C_{s}$ are in the interval $0$ $fm^{-1}<C_{s}\leq 9.6$
$fm^{-1}$ with $E_{0,\kappa }\neq -M$ when $C_{s}=0.$

Next, in order to establish the wave functions $F_{n\kappa }(r)$ of Eq. (8),
the relations (A7-A10) are used. Firstly, we calculate the first part of the
wave functions, 
\begin{equation}
\phi (x)=x^{\nu _{1}}(1-x)^{\kappa +1},\text{ }\nu _{1}>0.
\end{equation}%
The weight function takes the form%
\begin{equation}
\rho (x)=x^{2\nu _{1}}(1-x)^{2\kappa +1}.
\end{equation}%
which can generate the second part of the wave functions,%
\begin{equation}
y_{n}(x)\sim x^{-2\nu _{1}}(1-x)^{-\left( 2\kappa +1\right) }\frac{d^{n}}{%
dx^{n}}\left[ x^{n+2\nu _{1}}\left( 1-x\right) ^{n+2\kappa +1}\right]
\approx P_{n}^{(2\nu _{1},2\kappa +1)}(1-2x),
\end{equation}%
where $P_{n}^{(a,b)}(1-2x)$ is the orthogonal Jacobi polynomials [45,46].
Finally, the upper spinor component $F_{n\kappa }(x)$ for any arbitrary $%
\kappa $ can be obtained by means of Eq. (18) as%
\begin{equation*}
F_{n\kappa }(r)=\mathcal{N}_{n\kappa }e^{-2\nu _{1}\alpha r}(1-e^{-2\alpha
r})^{\kappa +1}P_{n}^{(2\nu _{1},2\kappa +1)}(1-2e^{-2\alpha r}),
\end{equation*}%
\begin{equation}
P_{n}^{(2\nu _{1},2\kappa +1)}(1-2e^{-2\alpha r})=\frac{\Gamma (n+2\nu
_{1}+1)}{\Gamma (2\nu _{1}+1)n!}%
\begin{array}{c}
_{2}F_{1}%
\end{array}%
\left( -n,n+2\left( \nu _{1}+\kappa +1\right) ;1+2\nu _{1};e^{-2\alpha
r}\right) ,
\end{equation}%
where the normalization constants $\mathcal{N}_{n\kappa }$ are calculated in
Appendix B.

The derivative relation of the hypergeometric function, 
\begin{equation*}
\frac{d}{dx}\left[ 
\begin{array}{c}
_{2}F_{1}%
\end{array}%
\left( a;b;c;x\right) \right] =\left( \frac{ab}{c}\right) 
\begin{array}{c}
_{2}F_{1}%
\end{array}%
\left( a+1;b+1;c+1;x\right) ,
\end{equation*}%
is usually used to calculate the corresponding lower-component $G_{n\kappa
}(r)$ by means of Eq. (9): 
\begin{equation*}
G_{n\kappa }(r)=\frac{\mathcal{N}_{n\kappa }}{Mc^{2}+E_{n\kappa }-C_{s}}%
\left( \frac{2\alpha (\kappa +1)e^{-2\alpha r}}{1-e^{-2\alpha r}}-2\alpha
\nu _{1}+\frac{\kappa }{r}\right) F_{n\kappa }(r)
\end{equation*}%
\begin{equation*}
+\mathcal{N}_{n\kappa }\frac{2n\alpha \left( n+2\nu _{1}+2\kappa +2\right) }{%
\left( Mc^{2}+E_{n\kappa }-C_{s}\right) \left( 1+2\nu _{1}\right) }%
(1-e^{-2\alpha r})^{\kappa +1}\left( e^{-2\alpha r}\right) ^{\nu _{1}+1}
\end{equation*}%
\begin{equation}
\times 
\begin{array}{c}
_{2}F_{1}%
\end{array}%
\left( -n+1;n+2\left( \nu _{1}+\kappa +1\right) +1;2\left( 1+\nu _{1}\right)
;e^{-2\alpha r}\right) .
\end{equation}%
The hypergeometric series $%
\begin{array}{c}
_{2}F_{1}%
\end{array}%
\left( -n+1;n+2\left( \nu _{1}+\kappa +1\right) +1;2\left( 1+\nu _{1}\right)
;e^{-\alpha r}\right) $ is terminated for $n=0$ and thus does not diverge
for all values of real parameters $\nu _{1}$ and $\kappa +1.$ In
relativistic units $\hbar =c=1,$ $E_{n\kappa }\neq -M,$ i.e., negative
energy states are forbidden, when $C_{s}=0$ in which the positive energy
solution of Eq. (25) is required. Therefore, the negative solution is not
desirable, see Fig. 2b.$.$

In Fig. 3, we plot the upper and lower spinor wave functions of ground $%
0p_{1/2}$ and first excited $1p_{1/2}$ states for (a) particle and (b)
antiparticle with $\kappa =1.$ The choices of parameters $M=5.0$ $fm^{-1},$ C%
$_{s}=4.9$ $fm^{-1},$ $\alpha =0.1$ $fm^{-1}$ and $A=1$ are used. In case of
the particle (positive energy), the upper and lower spinor wave functions of
the ground (first excited) state are found to be similar in shape. It is
noted that the amplitude of the upper wave function $F_{01}^{+}(r)$ ($%
F_{11}^{+}(r))$ is nearly three times larger than the lower wave function $%
G_{01}^{+}(r)$ ($G_{11}^{+}(r))$. Further, the amplitude of the ground state
wave function $F_{01}^{+}(r)$ is nearly two times larger than the first
excited state wave function $F_{11}^{+}(r)$. The range of the upper
component is wider than the lower component of the wave function and the
range of the first excited state is wider than the range of the ground state
wave function.

Let us now study the nonrelativistic case. Making the appropriate changes: $%
\nu _{1}^{2}\rightarrow \varepsilon _{nl}^{2}=-\left( 2mE_{nl}/4\alpha
^{2}\hbar ^{2}\right) ,$ $\omega _{1}\rightarrow \gamma _{1}=\left(
mA/\alpha \hbar ^{2}\right) $ and $\kappa (\kappa +1)\rightarrow l(l+1)$ in
Eqs. (16), (21)-(24) and (29)-(32) together with Table 1, we can easily
obtain the energy spectrum of the Schr\"{o}dinger equation for the Yukawa
potential model:%
\begin{equation}
\varepsilon _{nl}=\frac{\gamma _{1}}{2\left( n+l+1\right) }-\frac{\left(
n+l+1\right) }{2},
\end{equation}%
which can be explicitly expressed as%
\begin{equation}
E_{nl}=-\frac{\hbar ^{2}}{2m}\left[ \frac{mA}{\hbar ^{2}\left( n+l+1\right) }%
-\left( n+l+1\right) \alpha \right] ^{2},
\end{equation}%
and the radial wave functions: 
\begin{equation*}
R_{nl}(r)=r^{-1}u_{nl}(r)=\mathcal{N}_{nl}r^{-1}e^{-\epsilon
_{nl}r}(1-e^{-2\alpha r})^{l+1}P_{n}^{(2\varepsilon
_{nl},2l+1)}(1-2e^{-2\alpha r})
\end{equation*}%
\begin{equation}
=\mathcal{N}_{nl}\frac{\Gamma (n+2\varepsilon _{nl}+1)}{\Gamma (2\varepsilon
_{nl}+1)n!}e^{-\epsilon _{nl}r}(1-e^{-2\alpha r})^{l+1}%
\begin{array}{c}
_{2}F_{1}%
\end{array}%
\left( -n,n+2\left( \varepsilon _{nl}+l+1\right) ;1+2\varepsilon
_{nl};e^{-2\alpha r}\right) ,
\end{equation}%
where $\epsilon _{nl}=2\alpha \varepsilon _{nl}=\left[ \frac{mA}{\hbar
^{2}\left( n+l+1\right) }-\left( n+l+1\right) \alpha \right] $ and the
normalization constants $\mathcal{N}_{nl}$ are carried out in Appendix B.

\subsection{Pseudospin symmetry Dirac-Yukawa problem}

From Eq. (10), we can see that the energy eigenvalues depend mainly only on $%
n$ and $\widetilde{l},$ i.e., $E_{n\kappa }=E(n,\widetilde{l}(\widetilde{l}%
+1)).$ For $\widetilde{l}\neq 0,$ the states with $j=\widetilde{l}\pm 1/2$
are degenerate. This is a $SU(2)$ pseudospin symmetry. We impose $\Delta (r)$
as the Yukawa potential model given in (1):%
\begin{equation}
\Delta (r)=-\frac{\Delta _{0}}{r}e^{-\alpha r},\text{ }\Delta _{0}=A>0
\end{equation}%
leading to a Schr\"{o}dinger-like equation in the form:%
\begin{equation}
\left[ \frac{d^{2}}{dr^{2}}-\frac{\kappa \left( \kappa -1\right) }{r^{2}}%
-4\alpha ^{2}\nu _{2}^{2}+2\alpha \omega _{2}\frac{A}{r}e^{-\alpha r}\right]
G_{n\kappa }(r)=0,
\end{equation}%
with%
\begin{equation}
\nu _{2}^{2}=\frac{1}{4\alpha ^{2}\hbar ^{2}c^{2}}\left( Mc^{2}+E_{n\kappa
}\right) \left( Mc^{2}-E_{n\kappa }+C_{ps}\right) ,\text{ \ }\omega _{2}=%
\frac{1}{2\alpha \hbar ^{2}c^{2}}\left[ E_{n\kappa }-Mc^{2}-C_{ps}\right] A,
\end{equation}%
where $\kappa \left( \kappa -1\right) =\widetilde{l}(\widetilde{l}+1)$
satisfying Eq. (6)$.$ We follow the same procedures of solutions discussed
before to obtain a Dirac equation satisfying $G_{n\kappa }(r),$%
\begin{equation*}
\left\{ \frac{d^{2}}{dx^{2}}+\frac{(1-x)}{x(1-x)}\frac{d}{dx}\right.
\end{equation*}%
\begin{equation}
+\left. \frac{\left[ -\left( \nu _{2}^{2}+\omega _{2}\right) x^{2}+\left(
2\nu _{2}^{2}+\omega _{2}-\kappa \left( \kappa -1\right) \right) x-\nu
_{2}^{2}\right] }{x^{2}(1-x)^{2}}\right\} G_{n\kappa }(x)=0.
\end{equation}%
To avoid repetition in the solution of Eq. (40), a careful inspection for
the relationship between the present set of parameters $(\omega _{2},\nu
_{2}^{2})$ and the previous one $(\omega _{1},\nu _{1}^{2})$ tells us that
the negative energy solution for pseudospin symmetry, where $S(r)\sim -V(r),$
can be obtained directly from the spin symmetric solution by using the
following parameter mapping [40,47]: 
\begin{equation}
F_{n\kappa }(r)\leftrightarrow G_{n\kappa }(r),\text{ }\kappa \rightarrow
\kappa -1,\text{ }V(r)\rightarrow -V(r)\text{ (i.e., }A\rightarrow -A\text{)}%
,\text{ }E_{n\kappa }\rightarrow -E_{n\kappa }\text{ and }C_{s}\rightarrow
-C_{ps}.
\end{equation}%
Further, the constants in the case of pseudospin symmetry concept are listed
in Table 1. Applying the above transformations to Eqs. (21)-(24) leading to
the following pseudospin symmetric energy equation,%
\begin{equation}
\left( n+\kappa \right) ^{2}+2\left( n+\kappa \right) \nu _{2}=\omega _{2}.
\end{equation}%
Finally, with the aid of Eq. (39), Eq. (42) can be also expressed in terms
of the energy, 
\begin{equation*}
\sqrt{\left( Mc^{2}+E_{n\kappa }\right) \left( Mc^{2}-E_{n\kappa
}+C_{ps}\right) }+\alpha \hbar c\left( n+\kappa \right) =\frac{\left(
E_{n\kappa }-Mc^{2}-C_{ps}\right) A}{2\hbar c\left( n+\kappa \right) },\text{
}
\end{equation*}%
\begin{equation}
Mc^{2}>-E_{n\kappa }\text{ and }Mc^{2}+C_{ps}>E_{n\kappa },\text{ }%
n=0,1,2,\cdots ,
\end{equation}%
where $\kappa \neq -n.$ When $C_{ps}=0$ (exact pseudospin symmetry, $%
S(r)=-V(r)$ case$)$ then we require $-M<E_{n\kappa }<M.$The above pseudospin
energy equation can be rearranged in a quadratic form ($\hbar =c=1$):%
\begin{equation}
\left[ 1+\left( \frac{A}{N_{2}}\right) ^{2}\right] E_{n\kappa }^{2}-2\left[
T+\left( \frac{A}{N_{2}}\right) ^{2}U\right] E_{n\kappa }+\left( \frac{AU}{%
N_{2}}\right) ^{2}+\left( \frac{\alpha N_{2}}{2}\right) ^{2}-\left( M-\alpha
A\right) U=0,
\end{equation}%
where 
\begin{equation}
N_{2}=2\left( n+\kappa \right) ,\text{ }U=C_{ps}+M,\text{ }T=\frac{1}{2}%
(C_{ps}+\alpha A).
\end{equation}%
The above quadratic energy equation can be easily obtained by means of Eq.
(25) through making the replacements: $n+\kappa +1\rightarrow n+\kappa ,$ $%
A\rightarrow -A,$ $E_{n\kappa }\rightarrow -E_{n\kappa }$ and $%
C_{s}\rightarrow -C_{ps}.$ The two energy solutions of the quadratic
equation (44) can be obtained as 
\begin{equation}
E_{n\kappa }^{\pm }=\frac{(A^{2}U+TN_{2}^{2})\pm \sqrt{%
(A^{2}U+TN_{2}^{2})^{2}-(A^{2}+N_{2}^{2})\left[ \left( AU+\frac{1}{2}\alpha
N_{2}^{2}\right) ^{2}-MUN_{2}^{2}\right] }}{A^{2}+N_{2}^{2}},
\end{equation}%
where $(A^{2}U+TN_{2}^{2})>\sqrt{(A^{2}+N_{2}^{2})\left[ \left( AU+\frac{1}{2%
}\alpha N_{2}^{2}\right) ^{2}-MUN_{2}^{2}\right] }$ for distinct particle $%
E_{n\kappa }^{p}$ and anti-particle $E_{n\kappa }^{a}$ real bound state
energies.

On the basis of pseudospin symmetry, Fig. 4 shows the ground state hole
energy level of particle and antiparticle for different values of the
coupling constant $A$ and the screening parameter (range) of the potential $%
\alpha =0.01,0.02,0.05$ and $0.10$ \ for two degenerate partners ($%
n=0,\kappa =2$) and ($n=1,\kappa =-3$) labelled as ($0d_{3/2},1d_{5/2}$)
with $M=5.0$ $fm^{-1}$ and $C_{ps}=-5.0$ $fm^{-1}.$ In this exploratory
investigation, as it should be expected, for a given value of $\alpha $ the
bound state becomes slowly (sharply) deeper for particle (antiparticle),
i.e., it becomes more (less) attractive, on increasing the coupling constant 
$A$ (heavy nucleus). On increasing the value of the screening parameter $%
\alpha $, the particle (antiparticle) becomes more (less) attractive as $A$
increasing. Therefore, in the pseudospin symmetry case, the antiparticle is
less attractive to heavier nuclei while in the spin symmetry case the
particle is more attractive to heavier nuclei. In addition, increasing the
value of screening parameter $\alpha $ in both particle and antiparticle
leads to less attractive interaction with heavier nucleus.

Furthermore, Fig. 5 shows the variation of the ground state hole energy for
particle and antiparticle as a function of different values of pseudospin
constant $C_{ps}$ plotted for several values of the spin-orbit $\kappa =1,3$
and $5.$ We take the set of parameters $M=5.0$ $fm^{-1},$ $\alpha =0.1$ $%
fm^{-1}$ and $A=1.$ The results presented in Fig. 5a show that the energy
difference of particle between the states $\kappa =1,3$ and $5$ is almost
same although the values of spin constant $C_{ps}$ increases in the interval 
$-8$ $fm^{-1}\leq C_{ps}\leq 20$ $fm^{-1}$. With an increasing of $\kappa $
value, $E_{0\kappa }^{+}$ is seen to fan out toward the stronger negative
energy spectrum (more attractive) when $-20$ $fm^{-1}\leq C_{ps}\leq -13$ $%
fm^{-1}.$ The scattering states could be seen in the short interval $-12$ $%
fm^{-1}\leq C_{ps}\leq -8$ $fm^{-1}.$ Figure 5b shows that the energy
difference of antiparticle between the states $\kappa =1,3$ and $5$ is
almost same although the values of spin constant $C_{ps}$ increases in the
interval $-20$ $fm^{-1}\leq C_{ps}\leq -11$ $fm^{-1}$ (i.e., antiparticle
energy state is not sensitive to spin-orbit quantum number $\kappa )$. With
the $\kappa $ increasing, $E_{0\kappa }^{-}$ fan out toward the stronger
(deeper) negative energy spectrum when $-8$ $fm^{-1}\leq C_{ps}\leq 20$ $%
fm^{-1}.$ The scattering states could be seen in the short interval $-11$ $%
fm^{-1}\leq C_{ps}\leq -8$ $fm^{-1}.$ In the presence of pseudospin, we
consider the physical case where the energy has to be negative (i.e., $%
E_{n\kappa }\neq M$ when $C_{ps}=0$) which is simply the case of
antiparticle. As seen in Fig. 5b, the range of the allowed pseudospin
constant $C_{ps}$ falls in the range \ $-8$ $fm^{-1}<C_{ps}<20$ $fm^{-1}$ in
which $E_{n\kappa }^{-}$ is sensitive to the influence of $\kappa $ and also
energy results satisfy the condition $\left\vert E_{n\kappa }\right\vert <M.$
Consequently, we choose $C_{ps}=-5.0$ $fm^{-1}$ in the present numerical
calculations.

We present the essential procedures in calculating the wave functions. The
first part of the wave functions is%
\begin{equation}
\phi (x)=x^{\nu _{2}}(1-x)^{\kappa },\text{ }\nu _{1}>0
\end{equation}%
and the weight function is%
\begin{equation}
\rho (x)=x^{2\nu _{2}}(1-x)^{2\kappa -1}.
\end{equation}%
and this generates the second part of the wave functions,%
\begin{equation}
y_{n}(x)\sim x^{-2\nu _{2}}(1-x)^{-\left( 2\kappa -1\right) }\frac{d^{n}}{%
dx^{n}}\left[ x^{n+2\nu _{2}}\left( 1-x\right) ^{n+2\kappa -1}\right]
\approx P_{n}^{(2\nu _{2},2\kappa -1)}(1-2x).
\end{equation}%
Finally, the lower spinor component $G_{n\kappa }(x)$ for arbitrary $\kappa $
can be obtained by means of Eq. (18) as%
\begin{equation*}
G_{n\kappa }(r)=\mathcal{N}_{n\kappa }e^{-2\nu _{2}\alpha r}(1-e^{-2\alpha
r})^{\kappa }P_{n}^{(2\nu _{2},2\kappa -1)}(1-2e^{-2\alpha r}),
\end{equation*}%
\begin{equation}
P_{n}^{(2\nu _{2},2\kappa -1)}(1-2e^{-2\alpha r})=\frac{\Gamma (n+2\nu
_{2}+1)}{\Gamma (2\nu _{2}+1)n!}%
\begin{array}{c}
_{2}F_{1}%
\end{array}%
\left( -n,n+2\left( \nu _{2}+\kappa \right) ;1+2\nu _{2};e^{-2\alpha
r}\right) .
\end{equation}%
The upper-component $F_{n\kappa }(r)$ can be calculated from Eq. (11) as
follows 
\begin{equation*}
F_{n\kappa }(r)=\frac{\mathcal{N}_{n\kappa }}{\left( Mc^{2}-E_{n\kappa
}+C_{ps}\right) }\left( \frac{2\alpha \kappa e^{-2\alpha r}}{1-e^{-2\alpha r}%
}-2\alpha \nu _{2}-\frac{\kappa }{r}\right) G_{n\kappa }(r)
\end{equation*}%
\begin{equation*}
+\mathcal{N}_{n\kappa }\frac{2n\alpha \left( n+2\nu _{2}+2\kappa \right) }{%
\left( Mc^{2}-E_{n\kappa }+C_{ps}\right) \left( 1+2\nu _{2}\right) }%
(1-e^{-2\alpha r})^{\kappa }\left( e^{-2\alpha r}\right) ^{\nu _{2}+1}
\end{equation*}%
\begin{equation}
\times 
\begin{array}{c}
_{2}F_{1}%
\end{array}%
\left( -n+1;n+2\left( \nu _{2}+\kappa \right) +1;2\left( 1+\nu _{2}\right)
;e^{-2\alpha r}\right) .
\end{equation}%
In relativistic units $\hbar =c=1,$ $E_{n\kappa }\neq M,$ i.e., positive
energy states are forbidden, when $C_{ps}=0$ in which the negative
(antiparticle) energy solution of Eq. (43) is required. Therefore, the
positive solution is not desirable, see Fig. 5a$.$ The hypergeometric series 
$%
\begin{array}{c}
_{2}F_{1}%
\end{array}%
\left( -n+1;n+2\left( \nu _{2}+\kappa \right) +1;2\left( 1+\nu _{2}\right)
;e^{-\alpha r}\right) $ terminates for $n=0$ and thus does not diverge for
all values of real parameters $\nu _{2}$ and $\kappa .$

In Fig. 6, we plot the upper and lower spinor wave functions of ground $%
0d_{3/2}$ and first excited $1d_{3/2}$ states for (a) particle and (b)
antiparticle with $\kappa =2.$ The set of parameters $M=5.0$ $fm^{-1},$ C$%
_{ps}=-5.0$ $fm^{-1},$ $\alpha =0.1$ $fm^{-1}$ and $A=1$ are used.

\section{A Few Special Cases}

Let us study four special cases. At first, we study the nonrelativistic (Schr%
\"{o}dinger-Yukawa) case by setting $C_{s}=0$ (exact spin symmetry limit)
and making the changes $\kappa \rightarrow l,$ $M+E_{n\kappa }\rightarrow
2\mu /\hbar ^{2}$ and $M-E_{n\kappa }\rightarrow -E_{n\kappa }.$ Hence, from
Eq. (25), it follows that 
\begin{equation}
E_{nl}=-\frac{2\mu }{\hbar ^{2}}\left[ \frac{A_{0}}{2\left( n+l+1\right) }-%
\frac{\hbar ^{2}}{2\mu }\left( n+l+1\right) \alpha \right] ^{2},\text{ }%
n=0,1,2,\cdots ,\text{ }l=0,1,2,\cdots
\end{equation}%
Further, in the limit when $\alpha =0,$ the above result reduces to the
well-known spectrum for the nonrelativistic Coulombic field$,$ $%
E_{nl}^{(C)}=-\frac{\mu A_{0}^{2}}{2\hbar ^{2}\left( n+l+1\right) ^{2}}$
with a wave functions $R_{nl}(r)=r^{-1}\chi _{nl}(r)=r^{-1}e^{-\beta
r}r^{l+1}L_{n}^{\left( 2l+1\right) }(2\beta r)$ with $\beta =\frac{\mu A}{%
\hbar ^{2}\left( n+l+1\right) }>0,$ and $L_{n}^{s}(x)$ is the Laguarre
function [22].

Second, spin symmetry Dirac-Coulomb ($\alpha \rightarrow 0)$ case%
\begin{equation}
\sqrt{\left( Mc^{2}-E_{n\kappa }\right) \left( Mc^{2}+E_{n\kappa
}-C_{s}\right) }=\frac{\left( Mc^{2}+E_{n\kappa }-C_{s}\right) A}{2\hbar
c\left( n+\kappa +1\right) },\text{ }n=0,1,2,\cdots ,
\end{equation}%
The above energy equation can be rearranged in a quadratic form ($\hbar =c=1$%
):%
\begin{equation}
\left[ 1+\left( \frac{A}{N_{1}}\right) ^{2}\right] E_{n\kappa }^{2}-\left[
C_{s}+2\left( \frac{A}{N_{1}}\right) ^{2}W\right] E_{n\kappa }+\left( \frac{%
AW}{N_{1}}\right) ^{2}+MW=0,
\end{equation}%
and the two energy solutions of the above equation can be obtained as 
\begin{equation}
E_{n\kappa }^{\pm }=\frac{(2A^{2}W+C_{s}N_{1}^{2})\pm \sqrt{%
(2A^{2}W+C_{s}N_{1}^{2})^{2}-4W(A^{2}+N_{1}^{2})\left[ A^{2}W+MN_{1}^{2}%
\right] }}{2\left( A^{2}+N_{1}^{2}\right) },
\end{equation}%
In the limitation of pseudospin symmetry, the Dirac-Ykawa problem reduces to
Dirac-Coulomb problem when $\alpha \rightarrow 0$,%
\begin{equation}
\sqrt{\left( Mc^{2}+E_{n\kappa }\right) \left( Mc^{2}-E_{n\kappa
}+C_{ps}\right) }=\frac{\left( E_{n\kappa }-Mc^{2}-C_{ps}\right) A}{2\hbar
c\left( n+\kappa \right) },\text{ }n=0,1,2,\cdots ,
\end{equation}%
and it can be rearranged in a quadratic form ($\hbar =c=1$) as%
\begin{equation}
\left[ 1+\left( \frac{A}{N_{2}}\right) ^{2}\right] E_{n\kappa }^{2}-\left[
C_{ps}T+2\left( \frac{A}{N_{2}}\right) ^{2}U\right] E_{n\kappa }+\left( 
\frac{AU}{N_{2}}\right) ^{2}-MU=0.
\end{equation}%
Thus, the two energy solutions of the above equation can be readily obtained
as 
\begin{equation}
E_{n\kappa }^{\pm }=\frac{(2A^{2}U+C_{ps}N_{2}^{2})\pm \sqrt{%
(2A^{2}U+C_{ps}N_{2}^{2})^{2}-4U(A^{2}+N_{2}^{2})\left[ A^{2}U-MN_{2}^{2}%
\right] }}{2\left( A^{2}+N_{2}^{2}\right) },
\end{equation}%
Third, on the base of the exact spin symmetry $(C_{s}=0)$, the energy
equation for Dirac-Yukawa problem becomes (in units $\hbar =c=1$)%
\begin{equation}
\sqrt{M-E_{n\kappa }}+\alpha \left( n+\kappa +1\right) =\frac{A\sqrt{%
M+E_{n\kappa }}}{2\left( n+\kappa +1\right) },\text{ }\left\vert E_{n\kappa
}\right\vert <M,\text{ }n=0,1,2,\cdots ,
\end{equation}%
with two energy solutions of the quadratic equation (25) can be obtained as 
\begin{equation}
E_{n\kappa }^{\pm }=\frac{(\alpha N_{1}^{2}-2MA)A\pm \sqrt{(\alpha
N_{1}^{2}-2MA)^{2}A^{2}-(A^{2}+N_{1}^{2})\left[ \left( \alpha
N_{1}^{2}-2MA\right) ^{2}-\left( 2MN_{1}\right) ^{2}\right] }}{2\left(
A^{2}+N_{1}^{2}\right) },
\end{equation}%
where $N_{1}$ is defined in Eq. (27). From Eqs. (32) and (33), the upper and
lower wave functions are 
\begin{equation}
F_{n\kappa }(r)=\mathcal{N}_{n\kappa }\frac{\Gamma (n+2\gamma +1)}{\Gamma
(2\gamma +1)n!}e^{-2\gamma \alpha r}(1-e^{-2\alpha r})^{\kappa +1}%
\begin{array}{c}
_{2}F_{1}%
\end{array}%
\left( -n,n+2\left( \gamma +\kappa +1\right) ;1+2\gamma ;e^{-2\alpha
r}\right) ,
\end{equation}%
and%
\begin{equation*}
G_{n\kappa }(r)=\frac{\mathcal{N}_{n\kappa }}{M+E_{n\kappa }}\left( \frac{%
2\alpha (\kappa +1)e^{-2\alpha r}}{1-e^{-2\alpha r}}-2\alpha \gamma +\frac{%
\kappa }{r}\right) F_{n\kappa }(r)
\end{equation*}%
\begin{equation*}
+\mathcal{N}_{n\kappa }\frac{2n\alpha \left( n+2\gamma +2\kappa +2\right) }{%
\left( M+E_{n\kappa }\right) \left( 1+2\gamma \right) }(1-e^{-2\alpha
r})^{\kappa +1}\left( e^{-2\alpha r}\right) ^{\gamma +1}
\end{equation*}%
\begin{equation}
\times 
\begin{array}{c}
_{2}F_{1}%
\end{array}%
\left( -n+1;n+2\left( \gamma +\kappa +1\right) +1;2\left( 1+\gamma \right)
;e^{-\alpha r}\right) ,\text{ }M\neq -E_{n\kappa }
\end{equation}%
respectively, where $\gamma =\frac{1}{2\alpha }\sqrt{M^{2}-E_{n\kappa }^{2}}%
. $

In view of the exact pseudospin symmetry ($C_{ps}=0$), the energy equation
for Dirac-Yukawa problem reads ($\hbar =c=1$) 
\begin{equation}
\sqrt{M^{2}-E_{n\kappa }^{2}}+\alpha \left( n+\kappa \right) =\frac{\left(
E_{n\kappa }-M\right) A}{2\left( n+\kappa \right) },\text{ }n=0,1,2,\cdots ,
\end{equation}%
where $\left\vert E_{n\kappa }\right\vert <M$ and with two energy solutions: 
\begin{equation}
E_{n\kappa }^{\pm }=\frac{(\alpha N_{2}^{2}+2MA)A\pm \sqrt{(\alpha
N_{2}^{2}+2MA)^{2}A^{2}-(A^{2}+N_{2}^{2})\left[ \left( \alpha
N_{2}^{2}+2MA\right) ^{2}-\left( 2MN_{2}\right) ^{2}\right] }}{2\left(
A^{2}+N_{2}^{2}\right) },
\end{equation}%
The lower and upper-component wave functions are%
\begin{equation}
G_{n\kappa }(r)=\mathcal{N}_{n\kappa }\frac{\Gamma (n+2\gamma +1)}{\Gamma
(2\gamma +1)n!}e^{-2\gamma \alpha r}(1-e^{-2\alpha r})^{\kappa }%
\begin{array}{c}
_{2}F_{1}%
\end{array}%
\left( -n,n+2\left( \gamma +\kappa \right) ;1+2\gamma ;e^{-2\alpha r}\right)
,
\end{equation}%
and 
\begin{equation*}
F_{n\kappa }(r)=\frac{\mathcal{N}_{n\kappa }}{\left( Mc^{2}-E_{n\kappa
}\right) }\left( \frac{2\alpha \kappa e^{-2\alpha r}}{1-e^{-2\alpha r}}%
-2\alpha \gamma -\frac{\kappa }{r}\right) G_{n\kappa }(r)
\end{equation*}%
\begin{equation*}
+\mathcal{N}_{n\kappa }\frac{2n\alpha \left( n+2\gamma +2\kappa \right) }{%
\left( Mc^{2}-E_{n\kappa }\right) \left( 1+2\gamma \right) }(1-e^{-2\alpha
r})^{\kappa }\left( e^{-2\alpha r}\right) ^{\gamma +1}
\end{equation*}%
\begin{equation}
\times 
\begin{array}{c}
_{2}F_{1}%
\end{array}%
\left( -n+1;n+2\left( \gamma +\kappa \right) +1;2\left( 1+\gamma \right)
;e^{-\alpha r}\right) .
\end{equation}%
respectively.

Fourth, let us find the analytic solution of the Yukawa plus an added
centrifugal-like term, i.e., $V(r)=-\frac{A}{r}e^{-\alpha r}+\frac{D}{r^{2}}%
. $ The aim behind this choice is to remove the degeneracy of energy for
various states. In view of spin symmetry$,$ after inserting $\Sigma (r)=-%
\frac{A}{r}e^{-\alpha r}+\frac{D}{r^{2}}$ into Eq. (8), we obtain%
\begin{equation}
\left[ \frac{d^{2}}{dr^{2}}-\frac{\kappa ^{\prime }\left( \kappa ^{\prime
}+1\right) }{r^{2}}-4\alpha ^{2}\nu _{1}^{2}+2\alpha \omega _{1}\frac{A}{r}%
e^{-\alpha r}\right] F_{n\kappa }(r)=0,
\end{equation}%
and, hence, the energy spectrum can be obtained from Eq. (28) by making the
change $\kappa \rightarrow \kappa ^{\prime }$ as 
\begin{equation}
\sqrt{\left( M-E_{n\kappa }\right) \left( M+E_{n\kappa }-C_{s}\right) }%
+\alpha \left( n+\kappa ^{\prime }+1\right) =\frac{\left( M+E_{n\kappa
}-C_{s}\right) A}{2\left( n+\kappa ^{\prime }+1\right) },\text{ }\kappa
^{\prime }\neq -\left( n+1\right) ,
\end{equation}%
where%
\begin{equation}
\kappa ^{\prime }=-\frac{1}{2}+\frac{1}{2}\sqrt{\left( 2\kappa +1\right)
^{2}+4D\left( M+E_{n\kappa ^{\prime }}-C_{s}\right) }.
\end{equation}%
Further, the upper and lower wave functions can be obtained simply via Eqs.
(32) and (33) as 
\begin{equation}
F_{n\kappa }(r)=\mathcal{N}_{n\kappa }e^{-2\nu _{1}\alpha r}(1-e^{-2\alpha
r})^{\kappa ^{\prime }+1}%
\begin{array}{c}
_{2}F_{1}%
\end{array}%
\left( -n,n+2\left( \nu _{1}+\kappa ^{\prime }+1\right) ;1+2\nu
_{1};e^{-2\alpha r}\right) ,
\end{equation}%
and%
\begin{equation*}
G_{n\kappa }(r)=\frac{\mathcal{N}_{n\kappa }}{M+E_{n\kappa }-C_{s}}\left( 
\frac{2\alpha (\kappa ^{\prime }+1)e^{-2\alpha r}}{1-e^{-2\alpha r}}-2\alpha
\nu _{1}+\frac{\kappa ^{\prime }}{r}\right) F_{n\kappa }(r)
\end{equation*}%
\begin{equation*}
+\mathcal{N}_{n\kappa }\frac{2n\alpha \left( n+2\nu _{1}+2\kappa ^{\prime
}+2\right) }{\left( M+E_{n\kappa }-C_{s}\right) \left( 1+2\nu _{1}\right) }%
(1-e^{-2\alpha r})^{\kappa ^{\prime }+1}\left( e^{-2\alpha r}\right) ^{\nu
_{1}+1}
\end{equation*}%
\begin{equation}
\times 
\begin{array}{c}
_{2}F_{1}%
\end{array}%
\left( -n+1;n+2\left( \nu _{1}+\kappa ^{\prime }+1\right) +1;2\left( 1+\nu
_{1}\right) ;e^{-2\alpha r}\right) .
\end{equation}%
respectively. In view of pseudospin symmetry, we obtain the energy spectrum
from Eq. (46) by making the change $\kappa \rightarrow \kappa ^{\prime
\prime }$ as%
\begin{equation}
\sqrt{\left( M+E_{n\kappa }\right) \left( M-E_{n\kappa }+C_{ps}\right) }%
+\alpha \left( n+\kappa ^{\prime \prime }\right) =\frac{\left( E_{n\kappa
}-M-C_{ps}\right) A}{2\left( n+\kappa ^{\prime \prime }\right) },\text{ }%
\kappa ^{\prime \prime }=-n,\text{ }
\end{equation}%
where%
\begin{equation}
\kappa ^{\prime \prime }=\frac{1}{2}+\frac{1}{2}\sqrt{\left( 2\kappa
-1\right) ^{2}+4D\left( E_{n\kappa }-M-C_{ps}\right) }.
\end{equation}%
The lower and upper wave functions can be obtained as follows 
\begin{equation*}
G_{n\kappa }(r)=\mathcal{N}_{n\kappa }e^{-2\nu _{2}\alpha r}(1-e^{-2\alpha
r})^{\kappa ^{\prime \prime }}%
\begin{array}{c}
_{2}F_{1}%
\end{array}%
\left( -n,n+2\left( \nu _{2}+\kappa ^{\prime \prime }\right) ;1+2\nu
_{2};e^{-2\alpha r}\right) ,
\end{equation*}%
and%
\begin{equation*}
F_{n\kappa }(r)=\frac{\mathcal{N}_{n\kappa }}{\left( M-E_{n\kappa
}+C_{ps}\right) }\left( \frac{2\alpha \kappa ^{\prime \prime }e^{-2\alpha r}%
}{1-e^{-2\alpha r}}-2\alpha \nu _{2}-\frac{\kappa ^{\prime \prime }}{r}%
\right) G_{n\kappa }(r)
\end{equation*}%
\begin{equation*}
+\mathcal{N}_{n\kappa }\frac{2n\alpha \left( n+2\nu _{2}+2\kappa ^{\prime
\prime }\right) }{\left( M-E_{n\kappa }+C_{ps}\right) \left( 1+2\nu
_{2}\right) }(1-e^{-2\alpha r})^{\kappa ^{\prime \prime }}\left( e^{-2\alpha
r}\right) ^{\nu _{2}+1}
\end{equation*}%
\begin{equation}
\times 
\begin{array}{c}
_{2}F_{1}%
\end{array}%
\left( -n+1;n+2\left( \nu _{2}+\kappa ^{\prime \prime }\right) +1;2\left(
1+\nu _{2}\right) ;e^{-2\alpha r}\right) .
\end{equation}%
Fifth, the exact Dirac-Coulomb problem ($\alpha =0$) has the following
energy equations:\tablenotemark[1]%
\tablenotetext[1]{The bound state
solutions are exact for the case $\alpha=0$ since it lies in the short
screening range.}%
\begin{equation}
4N_{ps}^{2}\left( M+E_{n\kappa }\right) =\left( M-E_{n\kappa }\right) A^{2},%
\text{ }N_{ps}=n+\kappa ,\text{ }n=0,1,2,\cdots ,
\end{equation}%
and%
\begin{equation}
4N_{s}^{2}\left( M-E_{n\kappa }\right) =A^{2}(M+E_{n\kappa }),\text{ }%
N_{s}=n+\kappa +1,\text{ }n=0,1,2,\cdots ,
\end{equation}%
in the limitation of the exact pseudospin ($C_{ps}=0$) and spin ($C_{s}=0$)
symmetry, respectively. Obviously, in making the following changes $%
E_{n\kappa }\rightarrow -E_{n\kappa }$ and $\kappa \rightarrow \kappa -1,$
one can easily switch off from spin symmetry, Eq. (76), into pseudospin
symmetry, Eq. (75). Furthermore, Eqs. (75) and (76) are identical to Eqs.
(37) and (47) of Ref. [48] (if one sets $A=0=C$ and $B\rightarrow A$). They
are also identical to Eqs. (40) and (52) of Ref. [49].

\section{Numerical Results}

From Eq. (35), for small potential strength parameter $A=\sqrt{2}$, we
calculate some numerical values of the bound state energies of the Schr\"{o}%
dinger equation with the Yukawa potential for various values of quantum
numbers $n,$ $l$ and screening parameter $\alpha $ ($\hbar =\mu =1$). Our
approximated results in Table 3 are compared with those of [16,17,19]
together with the results of [13] who solved the Schr\"{o}dinger equation
numerically. \ Our results are in good agreement for small $\alpha $ values
and becomes worse as $\alpha $ increases since the approximation used to
substitute the singular part of Yukawa potential, $r^{-1}$ and the
centrifugal term $r^{-2}$ are true for $\alpha r\ll 1.$ As the potential
strength parameter $A$ becomes larger, the numerical solution of the Schr%
\"{o}dinger equation is extremely difficult because the screened Coulomb
(Yukawa) potential is very deep and the wavefunction becomes very sharply
peaked near the origin. Because of the instability of the wave function the
energy eigenvalues become fairly inaccurate as the strength $A$ increases.

Based on the spin symmetry case, from Eq. (28), we can calculate some
relativistic particle $E_{n\kappa }^{+}$ and antiparticle $E_{n\kappa }^{-}$
bound state energies with values of parameters $\alpha =0.1$ $fm^{-1},$ $%
A=1.0,$ $M=5.0$ $fm^{-1}$ and $C_{s}=4.9$ $fm^{-1}$for various states with
quantum numbers $n$ and $l\left( \kappa <0,\kappa >0\right) $ in the units $%
\hbar =c=1$ are used. We present our results in Table 4. Hence, one can see
that there are degenerate eigenvalues of the spin partners within the
attractive scalar and repulsive vector Yukawa potentials. For example, the
Dirac spin doublet eigenstate partners are: $%
(1s_{1/2},0p_{3/2},2p_{3/2},1d_{5/2},3d_{5/2},2f_{7/2},4f_{7/2},3g_{9/2},4h_{11/2}), 
$ $(2s_{1/2},3p_{3/2},0d_{5/2},4d_{5/2},1f_{7/2},2g_{9/2},3h_{11/2}),$ and $%
(3s_{1/2},4p_{3/2},0f_{7/2},1g_{9/2},2h_{11/2},1p_{1/2},0d_{3/2})...$ etc.

Using Eq. (46), we can also calculate some relativistic particle $E_{n\kappa
}^{+}$ and antiparticle $E_{n\kappa }^{-}$ pseudospin symmetric bound state
energies with values of parameters $\alpha =0.1$ $fm^{-1},$ $A=1.0,$ $M=5.0$ 
$fm^{-1}$ and $C_{ps}=-5.0$ $fm^{-1}$for various states with quantum numbers 
$n$ and $\widetilde{l}\left( \kappa <0,\kappa >0\right) $ in the units $%
\hbar =c=1$. We present our results in Table 5. One can see that there are
degenerate eigenvalues of the pseudospin partners within the attractive
scalar and repulsive vector Yukawa potentials. As an example, the Dirac
pseudospin doublet eigenstate partners are: $(1p_{3/2},2s_{1/2},2d_{5/2}),$ $%
(1d_{5/2},2f_{7/2},0d_{3/2}),$ $(1f_{7/2},2g_{9/2},0f_{5/2},1d_{3/2}),$ $%
(1g_{9/2},2h_{11/2},0g_{7/2},1f_{5/2}),$ and $%
(1h_{11/2},0h_{9/2},1g_{7/2}),...$ etc. When $n=1,$ there is a finite number
of bound states where $\ 1\leq \widetilde{l}\leq 25$ $\left( \kappa =-25,%
\text{ }\kappa =24\right) $ and when $n=2,$ $1\leq \widetilde{l}\leq 26$ $%
\left( \kappa =-26,\text{ }\kappa =23\right) .$\ 

\section{Conclusion and Outlook}

We have investigated the approximate bound state solutions of the Dirac
equation for the screened Coulomb potential model with an arbitrary
spin-orbit $\kappa $- state under the conditions of the spin (pseudospin)
symmetry limitation in the framework of the shortcut of the NU method
including the usual approximation scheme to approximate the centrifugal
(pseudo-centrifugal) barrier term. By setting $\Sigma (r)$ ($\Delta (r)$) to
the spherically symmetric screened Coulomb potential model, we have derived
the Dirac bound state energy spectra and associated two-component spinor
wave functions for arbitrary spin-orbit $\kappa $ state that provides an
approximate solution to the spin- and pseudo-spin symmetry. We have also
discussed in detail how to choose the appropriate physical values for the
spin constant $C_{s}$ from Eq. (25). In the presence of spin symmetry, we
find that the appropriate value for $C_{s}$ falls in the range $-20$ $%
fm^{-1}\leq C_{s}\leq 8.8$ $fm^{-1}$ for positive energy part while $0$ $%
fm^{-1}<C_{s}\leq 9.6$ $fm^{-1}$ for the negative energy part. Thus, the
allowed values for the $C_{s}$ in considering the whole spectrum (particle
and antiparticle) falls in $4.9$ $fm^{-1}<C_{s}\leq 8.8$ $fm^{-1}$ after
neglecting the negative part of energy according to the requirement $%
E_{n,\kappa }\neq -M$. However, in the presence of pseudosymmetry, we find $%
C_{ps}$ value falls in the range $-8$ $fm^{-1}\leq C_{ps}\leq 20$ $fm^{-1}$
in the antiparticle energy spectrum demanding that $E_{n,\kappa }\neq M$.
The Yukawa interaction between an electron and heavy nucleus appears to be
less attractive as the value of $\alpha $ increases. The resulting solutions
of the wave functions are being expressed in terms of the Jacobi polynomials
(or hypergeometric functions). We have shown that the present spin symmetry
can be easily reduced to the non-relativistic solution when one inserts $%
S(r)=+V(r)$ $($i.e., $\Delta (r)=0$ or $C_{s}=0$). The non-relativistic
limits of our solution are obtained by imposing appropriate changes of
parameters $\kappa (\kappa +1)\rightarrow l(l+1)$ in the spin symmetry
limits. Furthermore, when $\alpha \rightarrow 0,$ our results can be reduced
to the well-known bound state solutions for the Coulomb potential model. We
must point out that the numerical calculations for eigenenergies of the
Dirac states involved in Eqs. (28) and (46) are sensitive to the choice of
the parameters $C_{s},$ $C_{ps},$ $\alpha ,$ $A$ and $M.$ The spin
(pseudospin) limit Dirac energy spectrum computed in Table 3 (Table 4) for
arbitrarily chosen set of parameters are in the form of valence (hole)
states. In order to remove the extra degeneracies in energy levels, we have
solved Dirac-Yukawa problem in the presence of spin and pseudospin symmetry
by adding a centrifugal-like term, i.e., $V(r)=-Ae^{-\alpha r}/r+D/r^{2}.$

Finally, the solution of the Dirac-Coulomb problem can be readily obtained
from our solutions by setting $\alpha =0.$ Hence, we can obtain expressions
for the exact energy eigenvalues for the exact spin and pseudospin
limitations. These exact solutions are identical to the ones found recently
in Refs. [48,49].

\acknowledgments The author thanks the two kind referees for their
enlightening suggestions which helped him to improve this work.

\appendix

\section{R\'{e}sum\'{e} of Parametric Generalization of NU Method}

We present the Nikiforov-Uvarov essential polynomials, root, eigenvalues and
wave functions expressed in terms of the constants $c_{i}$ ($i=1,$ $2,\cdots
,13)$ and $\xi _{j}$ (j$=0,$ $1,$ $2)$:

(i) Constants:%
\begin{equation*}
c_{4}=\frac{1}{2}\left( 1-c_{1}\right) ,\text{ }c_{5}=\frac{1}{2}\left(
c_{2}-2c_{3}\right) ,\text{ }c_{6}=c_{5}^{2}+\xi _{2},
\end{equation*}%
\begin{equation*}
\text{ }c_{7}=2c_{4}c_{5}-\xi _{1},\text{ }c_{8}=c_{4}^{2}+\xi _{0},\text{ }%
c_{9}=c_{3}\left( c_{7}+c_{3}c_{8}\right) +c_{6},
\end{equation*}%
\begin{equation*}
c_{10}=c_{1}+2c_{4}+2\sqrt{c_{8}}-1>-1,\text{ }c_{11}=1-c_{1}-2c_{4}+\frac{2%
}{c_{3}}\sqrt{c_{9}}>-1,
\end{equation*}%
\begin{equation}
c_{12}=c_{4}+\sqrt{c_{8}}>0,\text{ }c_{13}=-c_{4}+\frac{1}{c_{3}}\left( 
\sqrt{c_{9}}-c_{5}\right) >0,\text{ }c_{3}\neq 0.
\end{equation}%
(ii) Polynomials: 
\begin{equation}
\pi (x)=c_{4}+\sqrt{c_{8}}+c_{5}x-\left( \sqrt{c_{9}}+c_{3}\sqrt{c_{8}}%
\right) x,
\end{equation}%
\begin{equation}
k=-\left( c_{7}+2c_{3}c_{8}\right) -2\sqrt{c_{8}c_{9}},
\end{equation}%
\begin{equation}
\tau (x)=1+2\sqrt{c_{8}}-\left[ c_{2}+2\left( \sqrt{c_{9}}+c_{3}\sqrt{c_{8}}%
-c_{5}\right) \right] x,
\end{equation}%
\begin{equation}
\tau ^{\prime }(x)=-2c_{3}-2\left( \sqrt{c_{9}}+c_{3}\sqrt{c_{8}}\right) <0.
\end{equation}%
(iii) Energy equation:%
\begin{equation}
c_{2}n-\left( 2n+1\right) c_{5}+\left( 2n+1\right) \left( \sqrt{c_{9}}+c_{3}%
\sqrt{c_{8}}\right) +n(n-1)c_{3}+c_{7}+2c_{3}c_{8}+2\sqrt{c_{8}c_{9}}=0.
\end{equation}%
(iv) Wave functions:%
\begin{equation}
\rho (x)=x^{c_{10}}(1-c_{3}x)^{c_{11}},
\end{equation}%
\begin{equation}
\phi (x)=x^{c_{12}}(1-c_{3}x)^{c_{13}},\text{ }c_{12}>0,\text{ }c_{13}>0,
\end{equation}%
\begin{equation}
y_{n}(x)=P_{n}^{\left( c_{10},c_{11}\right) }(1-2c_{3}x),\text{ }c_{10}>-1,%
\text{ }c_{11}>-1,
\end{equation}%
\begin{equation*}
F_{n\kappa }(x)=\mathcal{N}_{n\kappa
}x^{c_{12}}(1-c_{3}x)^{c_{13}}P_{n}^{\left( c_{10},c_{11}\right)
}(1-2c_{3}x),
\end{equation*}%
\begin{equation}
=\mathcal{N}_{n\kappa }x^{c_{12}}(1-c_{3}x)^{c_{13}}%
\begin{array}{c}
_{2}F_{1}%
\end{array}%
\left( -n,1+c_{10}+c_{11}+n;c_{10}+1;c_{3}x\right) ,\text{ }x\in \left[
0,1/c_{3}\right] ,c_{3}\neq 0.
\end{equation}%
where $\mathcal{N}_{n\kappa }$ is a normalization constants. Further, $%
P_{n}^{\left( \mu ,\nu \right) }(x),$ $\mu >-1,\nu >-1$ and $x\in \lbrack
-1,1]$ are the Jacobi polynomials with%
\begin{equation}
P_{n}^{\left( \alpha ,\beta \right) }(1-2s)=\frac{\left( \alpha +1\right)
_{n}}{n!}%
\begin{array}{c}
_{2}F_{1}%
\end{array}%
\left( -n,1+\alpha +\beta +n;\alpha +1;s\right) .
\end{equation}

$\label{appendix}$

\section{Calculations of the Normalization Constants}

The normalization constant, $\mathcal{N}_{nl}$ can be determined in closed
form. We start by using the relation between the hypergeometric function and
the Jacobi polynomials (see formula (8.962.1) in [45]): 
\begin{equation*}
\begin{array}{c}
_{2}F_{1}%
\end{array}%
\left( -n,n+\nu +\mu +1;\nu +1;\frac{1-x}{2}\right) =\frac{n!}{\left( \nu
+1\right) _{n}}P_{n}^{\left( \nu ,\mu \right) }(x),
\end{equation*}%
\begin{equation}
\left( \nu +1\right) _{n}=\frac{\Gamma (n+\nu +1)}{\Gamma (\nu +1)},
\end{equation}%
to rewrite the wave functions in (32) as%
\begin{equation}
F_{n\kappa }(r)=\mathcal{N}_{n\kappa }\frac{n!\Gamma (2\nu _{1}+1)}{\Gamma
(n+2\nu _{1}+1)}e^{-2\nu _{1}\alpha r}(1-e^{-2\alpha r})^{\kappa
+1}P_{n}^{(2\nu _{1},2\kappa +1)}(1-2e^{-2\alpha r}).
\end{equation}%
From the normalization condition $\int_{0}^{\infty }\left[ u_{n,l}(r)\right]
^{2}dr=1$ and under the coordinate change $x=1-2e^{-2\alpha r},$ the
normalization constant in (B2) is given by%
\begin{equation}
\mathcal{N}_{n\kappa }^{-2}=\frac{1}{\alpha }\left[ \frac{n!\Gamma (2\nu
_{1}+1)}{\Gamma (n+2\nu _{1}+1)}\right] ^{2}\int_{-1}^{1}\left( \frac{1-x}{2}%
\right) ^{2\nu _{1}}\left( \frac{1+x}{2}\right) ^{2\kappa +1}\left( \frac{1+x%
}{2}\right) \left[ P_{n}^{(2\nu _{1},2\kappa +1)}(x)\right] ^{2}dx.
\end{equation}%
The calculation of this integral can be done by writing 
\begin{equation*}
\frac{1+x}{2}=1-\left( \frac{1-x}{2}\right) ,
\end{equation*}%
and using the following two integrals (see formula (7.391.5) in [45]):%
\begin{equation}
\int_{-1}^{1}\left( 1-x\right) ^{\nu -1}\left( 1+x\right) ^{\mu }\left[
P_{n}^{\left( \nu ,\mu \right) }(x)\right] ^{2}dx=2^{\nu +\mu }\frac{\Gamma
(n+\nu +1)\Gamma (n+\mu +1)}{n!\nu \Gamma (n+\nu +\mu +1)},
\end{equation}%
which is valid for $\func{Re}$($\nu )>0$ and $\func{Re}$($\mu )>-1$ and (see
formula (7.391.1) in [45]):%
\begin{equation}
\int_{-1}^{1}\left( 1-x\right) ^{\nu }\left( 1+x\right) ^{\mu }\left[
P_{n}^{\left( \nu ,\mu \right) }(x)\right] ^{2}dx=2^{\nu +\mu +1}\frac{%
\Gamma (n+\nu +1)\Gamma (n+\mu +1)}{n!\Gamma (n+\nu +\mu +1)(2n+\nu +\mu +1)}%
,
\end{equation}%
which is valid for $\func{Re}$($\nu )>-1,$ $\func{Re}$($\mu )>-1.$ Finally,
we have carried out relativistic and non-relativistic normalization
constants as%
\begin{equation}
\mathcal{N}_{n\kappa }=\frac{1}{\Gamma (2\nu _{1}+1)}\left[ \frac{\alpha \nu
_{1}(n+\nu _{1}+\kappa +1)}{(n+\kappa +1)}\frac{\Gamma (n+2\nu _{1}+1)\Gamma
(n+2\nu _{1}+2\kappa +2)}{n!\Gamma \left( n+2\kappa +2\right) }\right]
^{1/2},
\end{equation}%
and%
\begin{equation}
\mathcal{N}_{n\kappa }=\frac{1}{\Gamma (2\varepsilon _{nl}+1)}\left[ \frac{%
\alpha \varepsilon _{nl}(n+l+\varepsilon _{nl}+1)}{(n+l+1)}\frac{\Gamma
(n+2\varepsilon _{nl}+1)\Gamma (n+2\varepsilon _{nl}+2l+2)}{n!\Gamma \left(
n+2l+2\right) }\right] ^{1/2}.
\end{equation}%
respectively.

\newpage

{\normalsize 
}

\bigskip

\baselineskip= 2\baselineskip
\bigskip \newpage {\normalsize 
}

\baselineskip= 2\baselineskip
\bigskip \newpage \FRAME{ftbpFO}{0.0277in}{0.0277in}{0pt}{\Qct{Spin symmetry
Dirac ground valence state energy level for (a) a particle and (b)
anti-particle of mass $M=5.0$ $fm^{-1}$ as a function of the coupling
constant $A$ for several values of the range of the screened Coulomb
potential. }}{}{Figure 1}{}

\bigskip \FRAME{ftbpFO}{0.0277in}{0.0277in}{0pt}{\Qct{Dirac ground valence
state energy level for (a) a particle and (b) anti-particle of mass $M=5.0$ $%
fm^{-1}$ as a function of the spin symmetry constant $C_{s}$ for several
values of the spin-orbit $\protect\kappa .$}}{}{Figure 2}{}

\bigskip \FRAME{ftbpFO}{0.0277in}{0.0277in}{0pt}{\Qct{Spin symmetry Dirac
upper and lower spinor wave functions of ground $0p_{1/2}$ and first excited 
$1p_{1/2}$ states of (a) particle and (b) antiparticle with $\protect\kappa %
=1.$}}{}{Figure 3}{}

\bigskip

\FRAME{ftbpFO}{0.0277in}{0.0277in}{0pt}{\Qct{Pseudospin symmetry Dirac
ground hole state energy level for (a) a particle and (b) anti-particle of
mass $M=5.0$ $fm^{-1}$ as a function of the coupling constant $A$ for
several values of the range of the screened Coulomb potential. }}{}{Figure 4%
}{}

\bigskip \FRAME{ftbpFO}{0.0277in}{0.0277in}{0pt}{\Qct{Dirac ground hole
state energy level for (a) a particle and (b) anti-particle of mass $M=5.0$ $%
fm^{-1}$ as a function of the pseudospin symmetry constant $C_{ps}$ for
several values of the spin-orbit $\protect\kappa .$}}{}{Figure 5}{}

\bigskip \FRAME{ftbpFO}{0.0277in}{0.0277in}{0pt}{\Qct{Pseudospin symmetry
Dirac upper and lower spinor wave functions of ground $0d_{3/2}$ and first
excited $1d_{3/2}$ states of (a) particle and (b) antiparticle with $\protect%
\kappa =2.$}}{}{Figure 6}{}

\bigskip

\begin{table}[tbp]
\caption{Specific values of the constants in Dirac-Yukawa problem.}%
\begin{tabular}{lll}
\tableline Spin symmetry case &  & Pseudospin symmetry case \\ 
\tableline$c_{1}=1$ &  & $c_{1}=1$ \\ 
$c_{2}=1$ &  & $c_{2}=1$ \\ 
$c_{3}=1$ &  & $c_{3}=1$ \\ 
c$_{4}=0$ &  & c$_{4}=0$ \\ 
$c_{5}=-\frac{1}{2}$ &  & $c_{5}=-\frac{1}{2}$ \\ 
$c_{6}=\frac{1}{4}\left[ 1+4\left( \nu _{1}^{2}+\omega _{1}\right) \right] $
&  & $c_{6}=\frac{1}{4}\left[ 1+4\left( \nu _{2}^{2}+\omega _{2}\right) %
\right] $ \\ 
$c_{7}=-$2$\nu _{1}^{2}-\omega _{1}+\kappa \left( \kappa +1\right) $ &  & $%
c_{7}=-$2$\nu _{2}^{2}-\omega _{2}+\kappa \left( \kappa -1\right) $ \\ 
$c_{8}=\nu _{1}^{2}$ &  & $c_{8}=\nu _{2}^{2}$ \\ 
$c_{9}=\frac{1}{4}\left( 2\kappa +1\right) ^{2}$ &  & $c_{9}=\frac{1}{4}%
\left( 2\kappa -1\right) ^{2}$ \\ 
$c_{10}=2\nu _{1}$ &  & $c_{10}=2\nu _{2}$ \\ 
$c_{11}=2\kappa +1$ &  & $c_{11}=2\kappa -1$ \\ 
$c_{12}=\nu _{1}$ &  & $c_{12}=\nu _{2}$ \\ 
$c_{13}=\kappa +1$ &  & $c_{13}=\kappa $ \\ 
$\xi _{2}=\nu _{1}^{2}+\omega _{1}$ &  & $\xi _{2}=\nu _{2}^{2}+\omega _{2}$
\\ 
$\xi _{1}=2\nu _{1}^{2}+\omega _{1}-\kappa \left( \kappa +1\right) $ &  & $%
\xi _{1}=2\nu _{2}^{2}+\omega _{2}-\kappa \left( \kappa -1\right) $ \\ 
$\xi _{0}=\nu _{1}^{2}$ &  & $\xi _{0}=\nu _{2}^{2}$ \\ 
\tableline &  & 
\end{tabular}%
\end{table}

\bigskip \bigskip 
\begin{table}[tbp]
\caption{Spin symmetric Dirac-Yukawa ground valence energy spectrum (in
units of$\ fm^{-1})$ of the particle and antiparticle as a function of $%
C_{s} $ for various values of $\protect\kappa .$}%
\begin{tabular}{llllllllll}
\tableline$C_{s}$ & $E_{0,\kappa =1}^{+}$ & $E_{0,\kappa =3}^{+}$ & $%
E_{0,\kappa =5}^{+}$ & $E_{0,\kappa =7}^{+}$ & $E_{0,\kappa =9}^{-}$ & $%
C_{s} $ & $E_{0,\kappa =1}^{-}$ & $E_{0,\kappa =3}^{-}$ & $E_{0,\kappa
=5}^{-}$ \\ 
\tableline$-20$ & $3.328$ & $4.632$ & $4.880$ & $4.962$ & $4.992$ & $-20$ & $%
-24.999$ & $-24.995$ & $-24.988$ \\ 
$-15$ & $3.622$ & $4.707$ & $4.913$ & $4.977$ & $4.998$ & $-15$ & $-19.998$
& $-19.994$ & $-19.986$ \\ 
$-10$ & $3.916$ & $4.783$ & $4.943$ & $4.990$ & $5.000$ & $-10$ & $-14.998$
& $-14.992$ & $-14.982$ \\ 
$-5$ & $4.209$ & $4.857$ & $4.972$ & $4.999$ & $4.996$ & $-5$ & $-9.997$ & $%
-9.989$ & $-9.976$ \\ 
$0$ & $4.502$ & $4.929$ & $4.995$ & $4.997$ & $4.975$ & $0$ & $-4.996$ & $%
-4.984$ & $-4.964$ \\ 
$5$ & $4.792$ & $4.990$ & $4.993$ & $4.951$ & $4.883$ & $1$ & $-3.996$ & $%
-3.982$ & $-3.960$ \\ 
$6$ & $4.849$ & $4.998$ & $4.982$ & $4.921$ & $4.829$ & $4$ & $-0.993$ & $%
-0.974$ & $-0.940$ \\ 
$7$ & $4.905$ & $5.000$ & $4.958$ & $4.865$ & $4.726$ & $4.8$ & $-0.192$ & $%
-0.170$ & $-0.131$ \\ 
$8$ & $4.957$ & $4.988$ & $4.897$ & $4.722$ & $4.363$ & $4.9$ & $-0.092$ & $%
-0.069$ & $-0.030$ \\ 
$8.5$ & $4.980$ & $4.968$ & $4.818$ & $-$ & $-$ & $5$ & $0.007856$ & $0.032$
& $0.072$ \\ 
$8.8$ & $4.992$ & $4.942$ & $4.689$ & $-$ & $-$ & $6$ & $1.010$ & $1.039$ & $%
1.090$ \\ 
$9$ & $4.998$ & $4.910$ & $-$ & $-$ & $-$ & $7$ & $2.013$ & $2.053$ & $2.121$
\\ 
$9.5$ & $4.987$ & $-$ & $-$ & $-$ & $-$ & $8$ & $3.019$ & $3.079$ & $3.188$
\\ 
$10$ & $-$ & $-$ & $-$ & $-$ & $-$ & $9$ & $4.038$ & $4.173$ & $-$ \\ 
$10.5$ & $-$ & $-$ & $-$ & $-$ & $-$ & $9.5$ & $4.577$ & $-$ & $-$ \\ 
$10.6$ & $5.498$ & $-$ & $-$ & $-$ & $-$ & $9.6$ & $4.702$ & $-$ & $-$ \\ 
$10.7$ & $5.623$ & $-$ & $-$ & $-$ & $-$ & $9.7$ & $-$ & $-$ & $-$ \\ 
$10.8$ & $5.737$ & $-$ & $-$ & $-$ & $-$ & $10.5$ & $-$ & $-$ & $-$ \\ 
$10.9$ & $5.846$ & $-$ & $-$ & $-$ & $-$ & $10.6$ & $5.232$ & $-$ & $-$ \\ 
$11$ & $5.953$ & $5.754$ & $-$ & $-$ & $-$ & $11$ & $5.200$ & $5.360$ & $-$
\\ 
$11.1$ & $6.058$ & $5.899$ & $-$ & $-$ & $-$ & $11.3$ & $5.205$ & $5.272$ & $%
-$ \\ 
$11.3$ & $6.266$ & $6.147$ & $-$ & $-$ & $-$ & $11.4$ & $5.208$ & $5.258$ & $%
5.511$ \\ 
$11.4$ & $6.368$ & $6.262$ & $5.998$ & $-$ & $-$ & $11.5$ & $5.212$ & $5.247$
& $5.450$ \\ 
$12$ & $6.979$ & $6.912$ & $6.786$ & $6.562$ & $-$ & $12$ & $5.233$ & $5.218$
& $5.327$ \\ 
$13$ & $7.986$ & $7.944$ & $7.870$ & $7.759$ & $7.600$ & $15$ & $5.396$ & $%
5.208$ & $5.208$ \\ 
$14$ & $8.990$ & $8.959$ & $8.905$ & $8.828$ & $8.724$ & $20$ & $5.686$ & $%
5.268$ & $5.205$ \\ 
$15$ & $9.992$ & $9.967$ & $9.925$ & $9.866$ & $9.787$ &  &  &  &  \\ 
$20$ & $14.996$ & $14.984$ & $14.964$ & $14.935$ & $14.898$ &  &  &  &  \\ 
\tableline &  &  &  &  &  &  &  &  & 
\end{tabular}%
\end{table}

\bigskip

\begin{table}[tbp]
\caption{Nonrelativistic bound state energies of the Yukawa potential in
units of $\hbar =m=1.$ For comparison, we set $A=\protect\sqrt{2}$ and $%
\protect\alpha =gA.$}%
\begin{tabular}{llllllllll}
\tableline State & $n$ & $l$ & $g$ & NU (Present) & [13] (Numerical) & [23]
(AIM) & [21] (SUSY) & [16] & [17] \\ 
\tableline$1s$ & $0$ & $0$ & $0.002$ & $-0.996004$ & $-0.9960$ & $-0.996006$
& $-0.99601$ & $-0.99601$ & $-0.9960$ \\ 
&  &  & $0.005$ & $-0.990025$ & $-0.9900$ & $-0.990037$ & $-0.99004$ & $%
-0.99004$ & $-$ \\ 
&  &  & $0.01$ & $-0.9801$ & $-0.9801$ & $-0.980149$ & $-0.98015$ & $%
-0.98015 $ & $-0.9801$ \\ 
&  &  & $0.02$ & $-0.9604$ & $-0.9606$ & $-0.960592$ & $-0.96059$ & $%
-0.96059 $ & $-0.9606$ \\ 
&  &  & $0.025$ & $-0.950625$ & $-0.9509$ & $-0.950922$ & $-0.95092$ & $%
-0.95092$ & $-$ \\ 
&  &  & $0.05$ & $-0.9025$ & $-0.9036$ & $-0.903632$ & $-0.90363$ & $%
-0.90363 $ & $-0.9036$ \\ 
2s & $1$ & $0$ & $0.002$ & $-0.246016$ & $-0.2460$ & $-0.246023$ & $-0.24602$
& $-0.24602$ & $-0.2460$ \\ 
&  &  & $0.005$ & $-0.2401$ & $-0.2401$ & $-0.240148$ & $-0.24015$ & $%
-0.24015$ & $-$ \\ 
&  &  & $0.01$ & $-0.2304$ & $-0.2306$ & $-0.230586$ & $-0.23059$ & $%
-0.23058 $ & $-0.2306$ \\ 
&  &  & $0.02$ & $-0.2116$ & $-0.21230$ & $-0.212296$ & $-0.21230$ & $%
-0.21229$ & $-0.2124$ \\ 
&  &  & $0.025$ & $-0.2025$ & $-0.2036$ & $-0.203551$ & $-0.20355$ & $%
-0.20352$ & $-$ \\ 
&  &  & $0.05$ & $-0.16$ & $-0.1635$ & $-0.163542$ & $-0.16351$ & $-0.16325$
& $-0.1650$ \\ 
2p & $0$ & $1$ & $0.002$ & $-0.246016$ & $-0.2460$ & $-0.246019$ & $-0.24602$
& $-0.24602$ & $-0.2460$ \\ 
&  &  & $0.005$ & $-0.2401$ & $-0.2401$ & $-0.240123$ & $-0.24012$ & $%
-0.2412 $ & $-$ \\ 
&  &  & $0.01$ & $-0.2304$ & $-0.2305$ & $-0.230490$ & $-0.23049$ & $%
-0.23049 $ & $-0.2305$ \\ 
&  &  & $0.02$ & $-0.2116$ & $-0.2119$ & $-0.211926$ & $-0.21192$ & $%
-0.21193 $ & $-0.2120$ \\ 
&  &  & $0.025$ & $-0.2025$ & $-0.2030$ & $-0.202984$ & $-0.20299$ & $%
-0.20299$ & $-$ \\ 
&  &  & $0.05$ & $-0.16$ & $-0.1615$ & $-0.161480$ & $-0.16144$ & $-0.16155$
& $-0.1625$ \\ 
$3p$ & $1$ & $1$ & $0.002$ & $-0.107147$ & $-0.1072$ & $-0.107160$ & $%
-0.10716$ & $-0.10716$ & $-0.1072$ \\ 
&  &  & $0.005$ & $-0.101336$ & $-0.1014$ & $-0.101416$ & $-0.10142$ & $%
-0.10142$ & $-$ \\ 
&  &  & $0.01$ & $-0.092011$ & $-0.09231$ & $-0.092306$ & $-0.09231$ & $%
-0.09236$ & $-0.09236$ \\ 
&  &  & $0.02$ & $-0.074711$ & $-0.07570$ & $-0.075704$ & $-0.07570$ & $%
-0.07563$ & $-0.07611$ \\ 
&  &  & $0.025$ & $-0.066736$ & $-0.06816$ & $-0.068157$ & $-0.06814$ & $%
-0.06799$ & $-$ \\ 
&  &  & $0.05$ & $-0.033611$ & $-0.03712$ & $-0.037115$ & $-0.03739$ & $%
-0.03486$ & $-0.04236$ \\ 
$3d$ & $0$ & $2$ & $0.002$ & $-0.107147$ & $-0.1072$ & $-0.107152$ & $%
-0.10715$ & $-0.10715$ & $-0.1072$ \\ 
&  &  & $0.005$ & $-0.101336$ & $-0.1014$ & $-0.101368$ & $-0.1014$ & $%
-0.10137$ & $-$ \\ 
&  &  & $0.01$ & $-0.092011$ & $-0.09212$ & $-0.092122$ & $-0.09212$ & $%
-0.09212$ & $-0.09216$ \\ 
&  &  & $0.02$ & $-0.074711$ & $-0.07503$ & $-0.075030$ & $-0.07502$ & $%
-0.07504$ & $-0.07531$ \\ 
&  &  & $0.025$ & $-0.066736$ & $-0.06715$ & $-0.067146$ & $-0.06713$ & $%
-0.06718$ & $-$%
\end{tabular}%
\end{table}

\bigskip

\begin{table}[tbp]
\caption{Dirac valence bound state energies (in units of $\ fm^{-1})$ of the
spin symmetry Yukawa problem for various values of $n$ and $l.$}%
\begin{tabular}{lllllllll}
\tableline$l$ & $n$, $\kappa <0$ & ($l,$ $j$) & $E_{n,\kappa <0}^{-}$ & $%
E_{n,\kappa <0}^{+}$ & $n$, $\kappa >0$ & ($l,$ $j$) & $E_{n,\kappa >0}^{-}$
& $E_{n,\kappa >0}^{+}$ \\ 
\tableline$0$ & $0,-1$ & $0s_{1/2}$ & $-$ & $-$ &  &  &  &  \\ 
$0$ & $1,-1$ & $1s_{1/2}$ & $-0.098076$ & $4.05808$ &  &  &  &  \\ 
$0$ & $2,-1$ & $2s_{1/2}$ & $-0.0922956$ & $4.78641$ &  &  &  &  \\ 
$0$ & $3,-1$ & $3s_{1/2}$ & $-0.0826327$ & $4.94209$ &  &  &  &  \\ 
$0$ & $4,-1$ & $4s_{1/2}$ & $-0.0690436$ & $4.98904$ &  &  &  &  \\ 
$1$ & $0,-2$ & $0p_{3/2}$ & $-0.098076$ & $4.05808$ & $0,1$ & $0p_{1/2}$ & $%
-0.0922956$ & $4.78641$ \\ 
$1$ & $1,-2$ & $1p_{3/2}$ & $-$ & $-$ & $1,1$ & $1p_{1/2}$ & $-0.0826327$ & $%
4.94209$ \\ 
$1$ & $2,-2$ & $2p_{3/2}$ & $-0.098076$ & $4.05808$ & $2,1$ & $2p_{1/2}$ & $%
-0.0690436$ & $4.98904$ \\ 
$1$ & $3,-2$ & $3p_{3/2}$ & $-0.0922956$ & $4.78641$ & $3,1$ & $3p_{1/2}$ & $%
-0.0514655$ & $4.99998$ \\ 
$1$ & $4,-2$ & $4p_{3/2}$ & $-0.0826327$ & $4.94209$ & $4,1$ & $4p_{1/2}$ & $%
-0.0298154$ & $4.99395$ \\ 
$2$ & $0,-3$ & $0d_{5/2}$ & $-0.0922956$ & $4.78641$ & $0,2$ & $0d_{3/2}$ & $%
-0.0826327$ & $4.94209$ \\ 
$2$ & $1,-3$ & $1d_{5/2}$ & $-0.098076$ & $4.05808$ & $1,2$ & $1d_{3/2}$ & $%
-0.0690436$ & $4.98904$ \\ 
$2$ & $2,-3$ & $2d_{5/2}$ & $-$ & $-$ & $2,2$ & $2d_{3/2}$ & $-0.0514655$ & $%
4.99998$ \\ 
$2$ & $3,-3$ & $3d_{5/2}$ & $-0.098076$ & $4.05808$ & $3,2$ & $3d_{3/2}$ & $%
-0.0298154$ & $4.99395$ \\ 
$2$ & $4,-3$ & $4d_{5/2}$ & $-0.0922956$ & $4.78641$ & $4,2$ & $4d_{3/2}$ & $%
-0.0039874$ & $4.97759$ \\ 
$3$ & $0,-4$ & $0f_{7/2}$ & $-0.0826327$ & $4.94209$ & $0,3$ & $0f_{5/2}$ & $%
-0.0690436$ & $4.98904$ \\ 
$3$ & $1,-4$ & $1f_{7/2}$ & $-0.0922956$ & $4.78641$ & $1,3$ & $1f_{5/2}$ & $%
-0.0514655$ & $4.99998$ \\ 
$3$ & $2,-4$ & $2f_{7/2}$ & $-0.098076$ & $4.05808$ & $2,3$ & $2f_{5/2}$ & $%
-0.0298154$ & $4.99395$ \\ 
$3$ & $3,-4$ & $3f_{7/2}$ & $-$ & $-$ & $3,3$ & $3f_{5/2}$ & $-0.0039874$ & $%
4.97759$ \\ 
$3$ & $4,-4$ & $4f_{7/2}$ & $-0.098076$ & $4.05808$ & $4,3$ & $4f_{5/2}$ & $%
-0.0261492$ & $4.95362$ \\ 
$4$ & $0,-5$ & $0g_{9/2}$ & $-0.0690436$ & $4.98904$ & $0,4$ & $0g_{7/2}$ & $%
-0.0514655$ & $4.99998$ \\ 
$4$ & $1,-5$ & $1g_{9/2}$ & $-0.0826327$ & $4.94209$ & $1,4$ & $1g_{7/2}$ & $%
-0.0298154$ & $4.99395$ \\ 
$4$ & $2,-5$ & $2g_{9/2}$ & $-0.0922956$ & $4.78641$ & $2,4$ & $2g_{7/2}$ & $%
-0.0039874$ & $4.97759$ \\ 
$4$ & $3,-5$ & $3g_{9/2}$ & $-0.098076$ & $4.05808$ & $3,4$ & $3g_{7/2}$ & $%
-0.0261492$ & $4.95362$ \\ 
$4$ & $4,-5$ & $4g_{9/2}$ & $-$ & $-$ & $4,4$ & $4g_{7/2}$ & $-0.0607542$ & $%
4.92325$ \\ 
$5$ & $0,-6$ & $0h_{11/2}$ & $-0.0514655$ & $4.99998$ & $0,5$ & $0h_{9/2}$ & 
$-0.0298154$ & $4.99395$ \\ 
$5$ & $1,-6$ & $1h_{11/2}$ & $-0.0690436$ & $4.98904$ & 1,5 & $1h_{9/2}$ & $%
-0.0039874$ & $4.97759$ \\ 
$5$ & $2,-6$ & $2h_{11/2}$ & $-0.0826327$ & $4.94209$ & 2,5 & $2h_{9/2}$ & $%
-0.0261492$ & $4.95362$ \\ 
$5$ & $3,-6$ & $3h_{11/2}$ & $-0.0922956$ & $4.78641$ & 3,5 & $3h_{9/2}$ & $%
-0.0607542$ & $4.92325$%
\end{tabular}%
\end{table}

\bigskip

\begin{table}[tbp]
\caption{Dirac hole bound state energies (in units of $fm^{-1})$ of the
pseudospin symmetry Yukawa problem for various values of $n$ and $\widetilde{%
l}.$}%
\begin{tabular}{lllllllll}
\tableline$\widetilde{l}$ & $n$, $\kappa <0$ & ($l,$ $j$) & $E_{n,\kappa
<0}^{-}$ & $E_{n,\kappa <0}^{+}$ & $n-1$, $\kappa >0$ & ($l+2,$ $j+1$) & $%
E_{n,\kappa >0}^{-}$ & $E_{n,\kappa >0}^{+}$ \\ 
\tableline$1$ & $1,-1$ & $1s_{1/2}$ & - & - & $0,2$ & $0d_{3/2}$ & $%
-4.603587 $ & $-0.00817777$ \\ 
$2$ & $1,-2$ & $1p_{3/2}$ & $-3.917958$ & $-0.00204188$ & $0,3$ & $0f_{5/2}$
& $-4.749130$ & $-0.01843870$ \\ 
$3$ & $1,-3$ & $1d_{5/2}$ & $-4.603587$ & $-0.00817777$ & $0,4$ & $0g_{7/2}$
& $-4.791740$ & $-0.03287710$ \\ 
$4$ & $1,-4$ & $1f_{7/2}$ & $-4.749130$ & $-0.01843870$ & $0,5$ & $0h_{9/2}$
& $-4.799920$ & $-0.05156860$ \\ 
$5$ & $1,-5$ & $1g_{9/2}$ & $-4.791740$ & $-0.03287710$ & $0,6$ &  & $%
-4.791590$ & $-0.07461340$ \\ 
$6$ & $1,-6$ & $1h_{11/2}$ & $-4.799920$ & $-0.05156860$ & $0,7$ &  & $%
-4.772990$ & $-0.102140$ \\ 
$7$ & $1,-7$ &  & $-4.791590$ & $-0.07461340$ & $0,8$ &  & $-4.746630$ & $%
-0.134308$ \\ 
$8$ & $1,-8$ &  & $-4.772990$ & $-0.102140$ & $0,9$ &  & $-4.713610$ & $%
-0.171314$ \\ 
$9$ & $1,-9$ &  & $-4.746630$ & $-0.134308$ & $0,10$ &  & $-4.674380$ & $%
-0.213399$ \\ 
$10$ & $1,-10$ &  & $-4.713610$ & $-0.171314$ & $0,11$ &  & $-4.629040$ & $%
-0.260854$ \\ 
$15$ & $1,-15$ &  & $-4.454290$ & $-0.439464$ & $0,16$ &  & $-4.300520$ & $%
-0.594696$ \\ 
$20$ & $1,-20$ &  & $-3.993200$ & $-0.903412$ & $0,21$ &  & $-3.708850$ & $%
-1.188370$ \\ 
$25$ & $1,-25$ &  & $-2.938620$ & $-1.959250$ & $0,24$ &  & $-2.938620$ & $%
-1.959250$ \\ 
$1$ & $2,-1$ & $2s_{1/2}$ & $-3.917958$ & $-0.00204188$ & $1,2$ & $1d_{3/2}$
& $-4.749130$ & $-0.01843870$ \\ 
$2$ & $2,-2$ & $2p_{3/2}$ & $-$ & $-$ & $1,3$ & $1f_{5/2}$ & $-4.791740$ & $%
-0.03287710$ \\ 
$3$ & $2,-3$ & $2d_{5/2}$ & $-3.917958$ & $-0.00204188$ & $1,4$ & $1g_{7/2}$
& $-4.799920$ & $-0.05156860$ \\ 
$4$ & $2,-4$ & $2f_{7/2}$ & $-4.603587$ & $-0.00817777$ & $1,\ 5$ & $%
1h_{9/2} $ & $-4.791590$ & $-0.07461340$ \\ 
$5$ & $2,-5$ & $2g_{9/2}$ & $-4.749130$ & $-0.01843870$ & $1,6$ &  & $%
-4.772990$ & $-0.102140$ \\ 
$6$ & $2,-6$ & $2h_{11/2}$ & $-4.791740$ & $-0.03287710$ & $1,7$ &  & $%
-4.746630$ & $-0.134308$ \\ 
$7$ & $2,-7$ &  & $-4.799920$ & $-0.05156860$ & $1,8$ &  & $-4.713610$ & $%
-0.171314$ \\ 
$8$ & $2,-8$ &  & $-4.791590$ & $-0.07461340$ & $1,9$ &  & $-4.674380$ & $%
-0.213399$ \\ 
$9$ & $2,-9$ &  & $-4.772990$ & $-0.102140$ & $1,10$ &  & $-4.629040$ & $%
-0.260854$ \\ 
$10$ & $2,-10$ &  & $-4.746630$ & $-0.134308$ & $1,11$ &  & $-4.577470$ & $%
-0.314039$ \\ 
$15$ & $2,-15$ &  & $-4.519370$ & $-0.373394$ & $2,16$ &  & $-4.209880$ & $%
-0.685887$ \\ 
$20$ & $2,-20$ &  & $-4.108150$ & $-0.788067$ & $2,21$ &  & $-3.525210$ & $%
-1.372260$ \\ 
$25$ & $2,-25$ &  & $-3.291060$ & $-1.606630$ & $2,23$ &  & $-2.938620$ & $%
-1.95925$ \\ 
$26$ & $2,-26$ &  & $-2.938620$ & $-1.959250$ &  &  &  &  \\ 
\tableline &  &  &  &  &  &  &  &  \\ 
&  &  &  &  &  &  &  & 
\end{tabular}%
\end{table}

\bigskip

\bigskip

\bigskip

\bigskip

\bigskip

\bigskip

\bigskip

\bigskip

\bigskip


\begin{thebibliography}{99}
\bibitem{1} H. Yukawa, Proc. Phys. Math. Soc. 17 (1935) 48; H. Yukawa, Proc.
Phys. Math. Soc. 19 (1937) 1084.

\bibitem{2} R. Messina and H. Lowen, Phys. Rev. Lett. 91 (2003) 146101.

\bibitem{3} S. Kar and Y.K. Ho, Phys. Rev. A 75 (2007) 062509.

\bibitem{4} S.A. Khrapak, A.V. Ivlev, G.E. Morfill, S.K. Zhdanov and H.M.
Thomas, IEEE Transactions on Plasma Science 32 (2004) 555.

\bibitem{5} P.J. Siemens, Phys. Rev. C 1 (1970) 98.

\bibitem{6} S.J. Lee, H.H. Gan, E.D. Cooper and S. Das Gupta, Phys. Rev. C
40 (1989) 2585.

\bibitem{7} W. Greiner, Relativistic Quantum Mechanics: Wave Equations, 3rd
edn. (Springer, Berlin, 2000).

\bibitem{8} M.A. Preston and R.K. Bhaduri, Structure of the Nucleus
(Addison-Wesley, Reading, 1975). \ 

\bibitem{9} W. Frank, Nature 445 (2007) 156.

\bibitem{10} L.I. Schiff, Quantum Mechanics, 3rd edn. (McGraw-Hill,
Singapore, 1968), p.325.

\bibitem{11} O.A. Gomes, H. Chacham and J.R. Mohallem, Phys. Rev. 50 (1994)
228.

\bibitem{12} R. Dutt, A. Ray and P.P. Ray, Phys. Lett. A 83 (1981) 65; D.
Singh and Y.P. Varshni, Phys. Rev. A 28 (1983) 2606; H. de Meyer \textit{et
al}, J. Phys. A 18 (1985) 849.

\bibitem{13} F.J. Rogers, H.C. Jr Graboske and D.G. Hardwood, Phys. Rev. A 1
(1970) 1577.

\bibitem{14} C.S. Lai, Phys. Rev. A 26 (1982) 2245.

\bibitem{15} R. Dutt \textit{et al}, J. Phys. A18 (1985) 1379; C.S. Lam and
Y.P. Varshni, Phys. Rev. A 6 (1972) 1391.

\bibitem{16} A. Chatterjee, J. Phys. A: Math. Gen. 19 (1986) 3707.

\bibitem{17} C. Lee, Phys. Lett. A 267 (2000) 101.

\bibitem{18} A.D. Alhaidari, H. Bahlouli and M.S. Abdelmonem, J. Phys. A:
Math. Theor. 41 (2008) 032001.

\bibitem{19} M. Grant and C.S. Lai, Phys. Rev. A 20 (1979) 718.

\bibitem{20} J.D. Hirschfelder, J. Chem. Phys. 33 (1960) 1462 ; J.
Killingbeck, Phy. Lett. A 65 (1987) 87.

\bibitem{21} B. G\"{o}n\"{u}l, K. K\"{o}ksal and E. Bak\i r, Phys. Scr. 73
(2006) 279.

\bibitem{22} S.M. Ikhdair and R. Sever, J. Math. Chem. 41 (4) (2007) 329,
343.

\bibitem{23} M. Karakoc and I. Boztosun, Int. J. Mod. Phys. E 15 (6) (2006)
1253.

\bibitem{24} J. Broeckhove, F. Arickx, W. Vanroose and V.S. Vasilevsky, J.
Phys. A: Math. Gen. 37 (2004) 7769.

\bibitem{25} J. Aguilar and J.M. Combes, Commun. Math. Phys. 22 (1971) 269;
Y.K. Ho, Phys. Rep. 99 (1983) 1.

\bibitem{26} T.Y. Wu and W.Y. Pauchy Hwang, Relativistic Quantum Mechanics
and Quantum Fields (World Scientific, Singapore, 1991).

\bibitem{27} J.N. Ginocchio, Phys. Rep. 414 (2005) 165.

\bibitem{28} P.R. Page, T. Goldman and J.N. Ginocchio, Phys. Rev. Lett. 86
(2001) 204.

\bibitem{29} A. Arima, M. Harvey and K. Shimizu, Phys. Lett. B 30 (1969) 517.

\bibitem{30} K.T. Hecht and A. Adler, Nucl. Phys. A 137 (1969) 129.

\bibitem{31} J.N. Ginocchio and D.G. Madland, Phys. Rev. C 57 (1998) 1167.

\bibitem{32} A. Bohr, I. Hamarnoto and B.R. Motelson, Phys. Scr. 26 (1982)
267.

\bibitem{33} J. Dudek, W. Nazarewicz, Z. Szymanski and G.A. Leander, Phys.
Rev. Lett. 59 (1987) 1405.

\bibitem{34} D. Troltenier, C. bahri and J. P. Draayer, Nucl. Phys. A 586
(1995) 53.

\bibitem{35} S.M. Ikhdair, J. Math. Phys. 52 (2011) 052303; S.M. Ikhdair and
R. Sever, J. Math. Phys. 52 (2011); doi:10.1063/1.3671640.

\bibitem{36} S.M. Ikhdair and R. Sever, J. Phys. A: Math. Theor. 44 (2011)
355301; S.M. Ikhdair, C. Berkdemir and R. Sever, Appl. Math. Comput. 217
(22) (2011) 9019.

\bibitem{37} S.M. Ikhdair, J. Math. Phys. 51 (2010) 023525.

\bibitem{38} F. Dominguez-Adame and A. Rodriguez, Phys. Lett. A 198 (1995)
275.

\bibitem{39} A.F. Nikiforov and V.B. Uvarov, Special Functions of
Mathematical Physics (Birkh\"{a}user: Basel, 1988).

\bibitem{40} S.M. Ikhdair and R. Sever, Cent. Eur. J. Phys. 8 (2010) 652.

\bibitem{41} M.R. Setare and S. Haidari, Phys. Scr. 81 (2010) 065201.

\bibitem{42} R.L. Greene and C. Aldrich, Phys. Rev. A 14 (1976) 2363.

\bibitem{43} S.M. Ikhdair, Chem. Phys. 361 (2009) 9; S.M. Ikhdair, J.
Quantum Infor. Science 1 (2011) 73.

\bibitem{44} A.D. Alhaidari, H. Bahlouli and A. Al-Hasan, Phys. Lett. A 349
(2006) 87.

\bibitem{45} M. Abramowitz and I.A. Stegun, Handbook of Mathematical
Functions (Dover Publication, New York, 1972).

\bibitem{46} I.S. Gradshteyn and I.M. Ryzhik, Tables and integrals, series
and products (New York, Academic, 1969).

\bibitem{47} C. Berkdemir and Y.-F. Cheng, Phys. Scr. 79 (2009) 035003.

\bibitem{48} M. Hamzavi, A.A. Rajabi and H. Hassanabadi, Few-Body Syst. 48
(2010) 171.

\bibitem{49} O. Aydo\u{g}du and R. Sever, Ann. Phys.325 (2010) 373.
\end{thebibliography}
\end{document}